\documentclass[11pt,a4paper]{article}
\usepackage[dvipdfmx]{graphicx}
\usepackage{amsmath, amssymb}
\usepackage{color}
\usepackage{ulem}

\newcommand{\avr}[1]{\langle \,#1\,\rangle}
\newcommand{\vct}[1]{\boldsymbol{#1}}
\newcommand{\mtrix}[3]{\langle \,#1\,|\,#2\,|\,#3\,\rangle}

\begin{document}

\baselineskip=16pt

{
\noindent
\Large\textbf{The least-squares analysis of the moments\\
 of the charge distribution in the mean-field models
}
}

\vspace{\baselineskip}

\noindent
Toshio Suzuki\footnote[1]{kt.suzuki2th@gmail.com}

\vspace{0.5\baselineskip}

\noindent
\parbox[t]{14cm}{
Research Center for Electron Photon Science, Tohoku University,\\[2pt]
Sendai 982-0826, Japan
}

\begin{center}
\parbox[t]{14cm}{
\small

\baselineskip=12pt

The $n$th moment of the charge distribution is composed of not only the $m(\le n)$th
moments of the point proton distribution, but also the $m(\le (n-2))$th ones of
the point neutron distribution.
The experimental value observed through electromagnetic interaction
makes it possible to investigate the point proton and neutron distributions together.
Comparing the moments of the charge distribution in the mean-field models
with experimental values from electron scattering,
each value of the related moments of the point proton and neutron distributions
is estimated in $^{40}$Ca, $^{48}$Ca and $^{208}$Pb
by the least-squares analysis(LSA).
The LSA determines the value of the $m$th moment uniquely 
as a result of the sum rule for the coefficients of the regression equations.
The investigation of the high-order moments like the sixth one
provides detailed information on the nuclear models.
To obtain experimental values of such moments from electron scattering, however, 
a new way to analyze cross sections is required,
instead of the conventional Fourier-Bessel- and sum-of-Gaussians-methods.
}
 
\end{center}

\vspace{0.5\baselineskip}

\section{Introduction}\label{intro}

\begin{figure}[ht]
\centering{%
\includegraphics[scale=1]{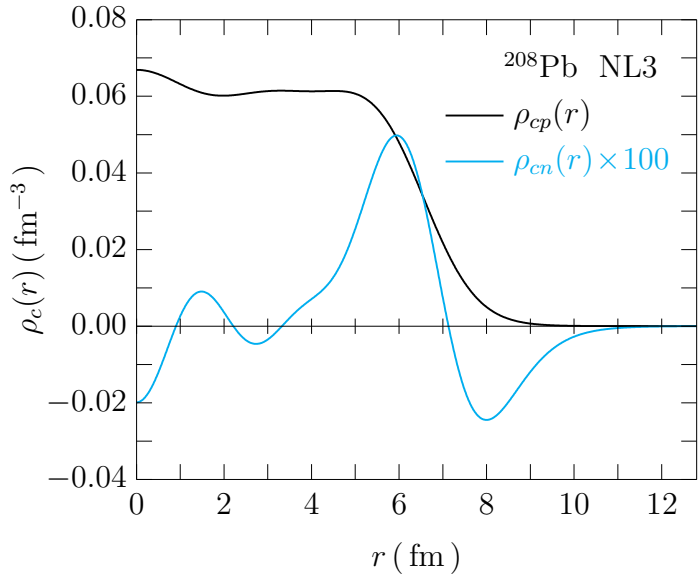}
}
\caption{The charge density($\rho_c(r)$)  of $^{208}$Pb
as a function of the nuclear coordinate($r$) calculated by the relativistic
mean-field model(NL3), taking into account the finite size of the nucleon.
The proton($\rho_{cp}(r)$) and the neutron($\rho_{cn}(r)$) charge densities are shown
by the black and blue curves, respectively. The former is almost the same as the total
charge density, $\rho_c(r)$. 
For details, see the text.
}
\label{cden_Pb208}
\end{figure}

Electron scattering is one of the best probes to investigate
the point proton distribution in nuclei
among various methods\cite{thi}.
Since the electromagnetic interaction and the reaction mechanism are well
understood theoretically\cite{bd},
the scattering process is completely separated from assumptions on the nuclear structure.
Moreover, the relationship between the nuclear charge distribution and the point proton
one is well defined\cite{deforest}.
The nuclear charge density is composed of the proton and neutron charge densities,
but is dominated by the former. The neutron density is less than about $1\%$
of the total density,
and its radial dependence oscillates to yield the integrated charge to be zero,
as shown in Fig. \ref{cden_Pb208}.
As a result, electron scattering has been used as a probe to explore
the point proton distribution throughout the periodic table\cite{vries},
and has not been employed so far to investigate
the point neutron distribution, which is another fundamental quantity to dominate
nuclear structure.
The neutron distribution has been studied with other probes mainly
through strong interaction which is dealt with rather model-dependently,
compared with electromagnetic interaction\cite{thi}.

Recently, it has been pointed out that the neutron charge density plays an appreciable role
in the $n$th(\,$n\ge 4$) moments of the nuclear charge distribution\cite{ks1},
because at the nuclear surface, contribution from the neutron charge density
to the total one becomes non-negligible, as expected from Fig. \ref{cden_Pb208}.
The $n$th moment of the nuclear charge distribution is shown to include the contribution
from the $m$($\le (n-2)$)th moment of the point neutron distribution,
in addition to the $m$($\le n$)th moment of the point proton distribution\cite{ks1}.
This fact implies that both the point proton and the point neutron distribution
are explored together, using the same experimental data through well-known
electromagnetic interaction.
It is not possible, however, to extract the neutron contribution
model-independently, because the $n$th(\,$n\ge 4$) moment has several components
whose values are not determined uniquely by experiment.
Strictly speaking, even the second moment of the charge density is not simply given by
the point proton distribution\cite{kss}.
Hence, in order to estimate the value of each component separately,
Ref.\cite{kss} has employed the least-squares analysis(LSA) for the values
of the moments predicted in the mean-field(MF) models, which
have been widely used to describe the fundamental properties of nuclei
for long time\cite{bm}.

The LSA used to analyze the moments is as follows. 
First, many MF models are arbitrarily chosen within the same model-framework
like the relativistic MF(RMF)-framework,
or the Skyrme-type MF(SMF)-framework
in the literature\cite{stone,nl3}.
In Ref.\cite{kss}, two sets composed of the $11$ RMF-models
and the $9$ SMF-ones
were prepared.
Next, the $n$th moment and its component, for example,
the fourth moment($Q^4_c$) of the charge distribution and the second moment
of the point neutron distribution($R^2_n$) in $^{48}$Ca, are calculated using each model.
Third, a pair of the two calculated values
are plotted on the $(R^2_n-Q^4_c)$-plane
as an element, ($R^2_{n,i}$, $Q^4_{c,i}$), by the model named $i$ in the set.
Fourth, the regression equation of the least-squares line(LSL)
for the elements is obtained, together with the
standard deviation and the correlation coefficient between two quantities.
Finally, the value of the component $R^2_n$ is determined by the cross
point(LSL-value) of the LSL with the line of the experimental value
of $Q^4_c$ observed through electron scattering.
Thus, the LSA is not a way to determine the experimental value of each component,
but to estimate the value accepted in the used model-framework.
Actually, Ref.\cite{kss}
has estimated the values of $R_n$ allowed in the RMF- and SMF-frameworks
for $^{40}$Ca, $^{48}$Ca and $^{208}$Pb 
within $1\%$ error\footnote[2]{Note that the
estimation of the error in Ref.\cite{kss} is different
from that in the present paper mentioned later.},
together with the ones of 
the root mean-square-radius(msr), $R_p$, of
the point proton distribution\footnote[3]{Note that the second moment is
frequently called the mean square radius, and its root is rms in the literature.},
using the experimental data available
at present\cite{vries, emrich}.

The present paper is an extension of Ref.\cite{kss}, where
the components of $Q^4_c$ in the MF models are explored.
In this paper, the sixth moments of the charge distribution($H^6_c$)
will be investigated. 
As far as the author knows, the structure of $H^6_c$ will be discussed
for the first time in nuclear physics.
Among the components of $H^6_c$,  the values of $R^2_n$
and the fourth moment of the point neutron distribution($Q^4_n$), 
and those of $R^2_p$, the fourth moment($Q^4_p$) and the
sixth moment of the point proton distribution($H^6_p$) will be estimated
in $^{40}$Ca, $^{48}$Ca and $^{208}$Pb.
The well defined LSLs 
between those moments and $H^6_c$ will be obtained
with the small standard
deviation and the correlation coefficients almost equal to $1$.
Unfortunately, however, there is no reliable experimental values of $H^6_c$ yet.
It will be shown in Appendix that the Fourier-Bessel(FB)-
and sum-of-Gaussians(SOG)-coefficients listed in Ref.\cite{vries}
is not sufficient for deducing the experimental values of $H^6_c$.
Hence, the obtained values by the FB- and SOG-methods will be called
the `experimental ones', but used for reference only in the discussions.

In this paper,
the structure of LSL will be also discussed in detail
so as to develop an understanding of the present analysis on $H^6_c$
and of previous one on $Q^4_c$\cite{kss}.
The LSL shows the linear relationship between the $n$th moment like $H^6_c$ and
its one component. Hence, it may not be obvious
whether or not the LSL-values can be used in the original expression of the $n$th moment
described as a function of the several components.
It will be shown that 
each value of the components in the function 
is uniquely given by the corresponding LSL-value in the model-framework,
and its function reproduces the experimental value of the $n$th moment.
This fact will be proved as a result of the sum rule for
the slope of the LSLs with respect to the $n$th moment.

In the next section,
the relativistic and non-relativistic expressions of $H^6_c$ will be derived
explicitly, following Ref.\cite{ks1}, where the equation of the $n$th
moment has been given.
The non-relativistic expression of the sixth moment is derived, in the same way
as for the second and the fourth moment\cite{kss}, by the Foldy-Wouthuysen
transformation of the four-component framework to the two-component one.
In \S \ref{ex}, the values of $H^6_c$ will be estimated by taking
a few MF-models as examples, so that the contribution of each components in $H^6_c$
would be understood quantitatively.
It will be shown that the contributions of the neutrons to $H^6_c$ are increased
more than to $Q^4_c$.
In \S \ref{lsa}, the LSA of the relationship between
various components and $H^6_c$ will be performed in the same way as
in Ref.\cite{kss} for $Q^4_c$.
Since the experimental values of $H^6_c$ are uncertain, the values obtained
from the SOG-method will be used for reference.
Following the results of \S \ref{lsa} and those of Ref.\cite{kss},
the structure of LSL
will be discussed in \S \ref{dis}.
The final section will be devoted to a brief summary.

\section{The relativistic and non-relativistic expressions of the sixth moment}\label{mom}

Neglecting the center of mass correction,
the sixth moment of the nuclear charge density is described as
\begin{equation}
H^6_c=\avr{r^6}_c=\sum_{\tau=p,n}\avr{r^6}_{c\tau}, \label{h}
\end{equation}
where $\avr{r^6}_{c\tau}$ is given by
\begin{equation}
\avr{r^6}_{c\tau}=\frac{1}{Z}\int d^3r r^6 \rho_{c\tau}(r),\label{hh}
\end{equation}
with the proton charge density, $\rho_{cp}(r)$, and the neutron charge density, $\rho_{cn}(r)$,
satisfying  $\int d^3r \rho_{cp}(r) =Z$, the number of the protons in the nucleus,
and $\int d^3r \rho_{cn}(r)=0$, respectively. Following Refs.\cite{ks1,kss},
it is convenient to use the Fourier component of $\rho_{c\tau}(r)$ for
calculations of $H^6_c$,
\begin{equation}
\tilde{\rho}_{c\tau}(q)=\int d^3r \exp(i\vct{q}\cdot\vct{r})\rho_{c\tau}(r)
 =\int d^3rj_0(qr)\rho_{c\tau}(r).
\end{equation}
Then, Eq.(\ref{hh}) is written as  
\begin{equation}
\avr{r^6}_{c\tau}=\frac{1}{Z}\left(-\vct{\nabla}^2_{\vct{q}}\right)^3
 \left(\tilde{\rho}_{G\tau}(q)+ \tilde{\rho}_{F\tau}(q)\right)|_{\vct{q}=0},\label{m}
\end{equation}
where
\begin{equation}
\tilde{\rho}_{G\tau}(q)=\int d^3r j_0(qr)G_{E\tau}(q^2)\rho_\tau(r)\,,\quad
\tilde{\rho}_{F\tau}(q)=\int d^3r j_0(qr)F_{2\tau}(q^2)W_\tau(r).
\end{equation}
In the above equations, $G_{E\tau}(q^2)$ and $F_{2\tau}(q^2)$ denote the Sachs and Pauli
form factors of the nucleon, and $\rho_\tau(r)$ and $W_\tau(r)$ stand for the relativistic
point nucleon distribution and the spin-orbit density, respectively.
The explicit forms of the nucleon form factors are provided in Ref.\cite{kss} as,
\begin{align}
G_{Ep}(q^2)&= \frac{1}{(1+r_p^2q^2/12)^2},
 \qquad F_{2p}=\frac{G_{Ep}(q^2)}
 {1+q^2/4M^2}, \label{expff}\\[4pt]
G_{En}(q^2)&= \frac{1}{(1+r_+^2q^2/12)^2}- \frac{1}{(1+r_-^2q^2/12)^2},\qquad
 F_{2n}=\frac{G_{Ep}(q^2)-G_{En}(q^2)/\mu_n}{1+q^2/4M^2},\nonumber
\end{align}
with
\begin{equation}
  r_p=0.877\, \textrm{fm}, \qquad r_{\pm}^2=(0.830)^2
  \mp0.058 \, \textrm{fm}^2. \nonumber
\end{equation}
The relativistic expressions of $\rho_\tau(r)$ and $W_\tau(r)$ are written in the MF
models as\cite{ks1,kss},
\begin{align} 
\rho_\tau(r)&= \sum_{\alpha\in\tau} \frac{2j_\alpha+1}{4\pi r^2}
 \left(G_\alpha^2 + F_\alpha^2\right),\label{d}\\
W_\tau(r)&= \frac{\mu_\tau}{M}\sum_{\alpha\in\tau} \frac{2j_\alpha+1}{4\pi r^2}
 \frac{d}{dr}\left(\frac{M-M^*(r)}{M}G_\alpha F_\alpha
+ \frac{\kappa_\alpha +1}{2Mr}G_\alpha^2 -
\frac{\kappa_\alpha - 1}{2Mr}F_\alpha^2\right),
\label{so}
\end{align}
where $j_\alpha$ denotes the total angular momentum
of a single-particle, $\kappa_\alpha=(-1)^{j_\alpha-\ell_\alpha
 +1/2}(j_\alpha+1/2)$, and $\ell_\alpha$ the orbital angular momentum.
The function 
$G_\alpha(r)$ and $F_\alpha(r)$ stand for the radial parts of
the large and small components of the single-particle wave function,
respectively, with the normalization,
\begin{equation}
\int_0^\infty\! dr \left(G_\alpha^2 + F_\alpha^2\right)=1.\label{norm}
\end{equation}
The spin-orbit density is a relativistic correction due to the anomalous
magnetic moment of the nucleon, $\mu_\tau$ ($\mu_p=1.793$,  $\mu_n=-1.913$ ),
and its role is enhanced by the effective nucleon mass,
$M^\ast(r)=M+V_\sigma(r)$, $V_\sigma(r)$ being the $\sigma$ meson-exchange
potential which behaves in the same way as the nucleon mass, $M$,
in the equation of motion. The value of $M^\ast$ depends on the relativistic
MF models, but is nearly equal to $0.6M$.
Eq.(\ref{d}) satisfies $\int d^3r \, \rho_\tau (r) = Z$ for $\tau = p$,
and $N$, the number of the neutrons in the nucleus, for $\tau = n$, respectively,
while Eq.(\ref{so}) does $\int d^3r\, W_\tau(r)=0$,
as it should.

Using Eq.(\ref{expff}) to (\ref{so}), Eq.(\ref{m}) is calculated 
in the same  ways as in Ref.\cite{ks1},
\begin{align}
\frac{1}{Z}\left(-\vct{\nabla}^2_{\vct{q}}\right)^3\tilde{\rho}_{G_p}(q)|_{\vct{q}=0}&=
\avr{r^6}_p+7r^2_p\avr{r^4}_p+\frac{35}{2}r^4_p\avr{r^2}_p+\frac{35}{3}r^6_p\,,\nonumber\\
\frac{1}{N}\left(-\vct{\nabla}^2_{\vct{q}}\right)^3\tilde{\rho}_{G_n}(q)|_{\vct{q}=0}&=
7(r^2_+-r^2_-)\avr{r^4}_n+\frac{35}{2}(r^4_+-r^4_-)\avr{r^2}_n
 +\frac{35}{3}(r^6_+-r^6_-)\,,\nonumber\\
\frac{1}{Z}\left(-\vct{\nabla}^2_{\vct{q}}\right)^3\tilde{\rho}_{F_p}(q)|_{\vct{q}=0}&= 
\avr{r^6}_{W_p}+7\left(r^2_p+\frac{3}{2M^2}\right)\avr{r^4}_{W_p}+\frac{35}{2}
\left(r^4_p+\frac{2r^2_p}{M^2}+\frac{3}{M^4}\right)\avr{r^2}_{W_p}\,,\nonumber\\
\frac{1}{N}\left(-\vct{\nabla}^2_{\vct{q}}\right)^3\tilde{\rho}_{F_n}(q)|_{\vct{q}=0}&= 
\avr{r^6}_{W_n}+7\left(s_2+\frac{3}{2M^2}\right)\avr{r^4}_{W_n}+\frac{35}{2}
\left(s_4+\frac{2}{M^2}s_2+\frac{3}{M^4}\right)\avr{r^2}_{W_n}\,,\nonumber\\
 &\qquad (\, s_k=r^k_p-(r^k_+-r^k_-)/\mu_n\, ).\nonumber
\end{align}
In the right-hand sides of the above equations, we have used the same notations as
in Ref.\cite{ks1}.
Finally, the relativistic expression of the sixth moment is described
in terms of $H^6_{cn}$ and $H^6_{cn}$ of 
the proton and neutron charge distributions expressed,
respectively, as
\begin{align}
H^6_c&=H^6_{cp}-H^6_{cn}\,,\label{hc}\\
&H^6_{cp}=\avr{r^6}_{cp}=H^6_p+H_{4p}+H_{2p}+H_{W_p}+\frac{35}{3}r^6_p\,,\label{cp}\\
&H^6_{cn}=-\avr{r^6}_{cn}=H_{4n}+H_{2n}+H_{W_n}-\frac{35}{3}(r^6_+-r^6_-)\frac{N}{Z}\,,\label{cn}
\end{align}
where the components of $H^6_{cn}$ and $H^6_{cn}$ are given by
\begin{align}
H^6_p&=\avr{r^6}_p\,,\quad H_{4p}=7r^2_p\avr{r^4}_p\,,
 \quad H_{2p}=\frac{35}{2}r^4_p\avr{r^2}_p\,,\nonumber\\
H_{W_p}&=\avr{r^6}_{W_p}+7\left(r^2_p+\frac{3}{2M^2}\right)\avr{r^4}_{W_p}+\frac{35}{2}
\left(r^4_p+\frac{2r^2_p}{M^2}+\frac{3}{M^4}\right)\avr{r^2}_{W_p}\,,\nonumber\\
H_{4n}&=-7(r^2_+-r^2_-)\avr{r^4}_n\frac{N}{Z}\,,
 \quad H_{2n}=-\frac{35}{2}(r^4_+-r^4_-)\avr{r^2}_n\frac{N}{Z}\,,\nonumber\\
H_{W_n}&=-\left[\avr{r^6}_{W_n}+7\left(s_2+\frac{3}{2M^2}\right)\avr{r^4}_{W_n}+\frac{35}{2}
\left(s_4+\frac{2}{M^2}s_2+\frac{3}{M^4}\right)\avr{r^2}_{W_n}\right]\frac{N}{Z}\,.\nonumber
\end{align}
with
\begin{equation}
Q^4_\tau=\avr{r^4}_\tau\,,\quad R^2_\tau=\avr{r^2}_\tau\,.
\end{equation}
The notation $\avr{r^n}_{W_\tau}$ is defined as
\[
 \avr{r^n}_{W_\tau}=\frac{1}{N_\tau}\int d^3rr^nW_\tau(r),
\]
where $N_\tau$ represents $Z$ and $N$ for $\tau=p$ and $\tau=n$, respectively.

The non-relativistic expression of the sixth moment
which is equivalent to the above relativistic one up to $1/M^2$ may be obtained 
by the Foldy-Wouthuysen(F-W) transformation\cite{ks1}.
For the phenomenological non-relativistic models, however,
it is not possible to derive the consistent relativistic corrections to them,
since their original four-component frameworks are not known.
In the present paper, the free Dirac equation is employed as the nuclear part of the
relativistic Hamiltonian for the F-W transformation\cite{ks1,kss}.
Then, the non-relativistic expression of the sixth moment is obtained from
Eq.(\ref{hc}) by replacing the relativistic matrix elements $\avr{r^n}_\tau$
and $\avr{r^n}_{W_\tau}$ with $\avr{r^n}_{{\rm nr+r},\tau}$ and $\avr{r^n}_{{\rm nr},W_\tau}$,
respectively,
\begin{align}
\avr{r^n}_{{\rm nr+r},\tau}&=\avr{r^n}_{{\rm nr},\tau}
 +\frac{n(n+1)}{8M^2}\avr{r^{n-2}}_{{\rm nr},\tau}+\frac{1}{2\mu_\tau}\avr{r^n}_{{\rm nr},W_\tau}\,,\label{non}\\
\avr{r^n}_{{\rm nr},W_\tau}&=\frac{n}{N_\tau}\frac{\mu_\tau}{2M^2}
 \mtrix{0}{\sum_{k\in\tau}r^{n-2}_k\vct{\ell}_k\cdot\vct{\sigma}_k}{0}_{\rm nr}\,,\label{nonw}
\end{align}
where the subscript nr of the matrix elements in the right-hand sides indicates that
they should be calculated in the two-component framework of the wave functions
\footnote[4]{The last term in the parenthesis
of Eq.(28) in Ref.\cite{kss} is missed in the relativistic correction of Ref.\cite{ks1}.}.

\section{Components of the sixth moment of the charge density}\label{ex}

\begin{table}
 \hspace*{1cm}%
\begin{tabular}{|l||c||c|c|c|c|c|} \hline
\rule{0pt}{12pt} &
 $ H^6_c$  &
 $H^6_{cp}$ &
 $H^6_p$&
 $H_{4p}$ &
$H_{2p}$ &
$H_{W_p}$ 
\\ \hline
\rule{0pt}{12pt}%
NL3                    &          &           &            &            &           &                \\
$^{40}$Ca     & $4817.801$ & $ 5001.186$ & $ 3878.987$ & $ 990.211$ & $ 118.059$ & $8.620$  \\
$^{48}$Ca     & $4100.526$ & $ 4588.453$ & $ 3485.765$ & $ 958.790$ & $ 118.189$ & $20.400$   \\ 
$^{208}$Pb    & $52219.819$ & $55001.458$ & $48112.714 $ & $6006.507$ & $308.614$ & $568.314$   \\ \hline
\rule{0pt}{12pt}%
NL-SH                  &          &           &          &           &          &           \\
$^{40}$Ca     & $ 4616.612$ & $ 4797.328$ & $ 3703.400$ & $ 961.631$ & $ 116.690$ & $10.299$  \\
$^{48}$Ca     & $ 3971.062$ & $ 4447.191$ & $ 3362.497$ & $ 940.040$ & $ 117.417$ & $21.929$  \\
$^{208}$Pb    & $50819.698$ & $53558.038$ & $46757.840$ & $5915.257$ & $ 306.920$ & $572.712$  \\ \hline
\rule{0pt}{12pt}%
SLy4                   &          &           &          &           &          &            \\
$^{40}$Ca     & $ 5227.890$ & $ 5405.694$ & $ 4228.162$ & $ 1055.877$ & $ 121.964$ & $-5.617$  \\
$^{48}$Ca     & $ 4817.510$ & $ 5270.163$ & $ 4075.773$ & $ 1066.215$ & $ 124.215$ & $-1.350$  \\
$^{208}$Pb    & $54008.716$ & $56323.109$ & $49573.044$ & $ 6098.248$ & $ 309.751$ & $336.756$ \\ \hline
\end{tabular}
\caption{
The sixth moment of the nuclear charge distribution($H^6_c$) in $^{40}$Ca, $^{48}$Ca and $^{208}$Pb
in units of fm$^6$.
They are calculated by using parameters of the relativistic nuclear
models, NL3\cite{nl3} and NL-SH\cite{nlsh}, and of the non-relativistic one, SLy4\cite{sly4}.
The values of $H^6_{cp}$ show the contributions from the proton charge distribution to $H^6_c$,
and other numbers the values of the components of $H^6_{cp}$.
The neutron charge distribution($H^6_{cn}$), which contributes to $H^6_c$ as $H^6_{c}=H^6_{cp}-H^6_{cn}$,
are listed in Table \ref{table_hn}.   For the details, see the text.
}
 \label{table_hp}
\end{table}

\begin{table}
 \hspace*{3cm}%
\begin{tabular}{|l||c|c|c|c|} \hline
\rule{0pt}{12pt} & 
$H^6_{cn}$&
$H_{4n}$&
$H_{2n}$&
$H_{W_n}$
\\ \hline
\rule{0pt}{12pt}%
NL3        &          &           &            &             \\
$^{40}$Ca  & $ 183.385$ & $ 139.496$  &$ 30.987$ & $ 10.969$  \\
$^{48}$Ca  & $ 487.927$ & $ 264.221$ & $ 50.887$ & $170.114$  \\ 
$^{208}$Pb & $2781.639$ & $1742.852$ & $141.626$ & $894.194$  \\ \hline
\rule{0pt}{12pt}%
NL-SH      &          &           &          &            \\
$^{40}$Ca  & $ 180.717$ & $ 135.540$ & $ 30.643$ & $ 12.602$  \\
$^{48}$Ca  & $ 476.129$ & $ 255.798$ & $ 50.274$ & $167.352$  \\
$^{208}$Pb & $2738.340$ & $1701.125$ & $140.167$ & $894.081$ \\ \hline
\rule{0pt}{12pt}%
SLy4       &          &           &          &            \\
$^{40}$Ca  & $ 177.805$ & $ 148.495$ & $ 31.941$ & $ -4.562$  \\
$^{48}$Ca  & $ 452.653$ & $ 267.827$ & $ 51.189$ & $130.933$ \\
$^{208}$Pb & $2314.393$ & $1611.226$ & $136.008$ & $564.191$ \\ \hline
\end{tabular}
\caption{The sixth moment of
the neutron charge distribution($H^6_{cn}$)
in $^{40}$Ca, $^{48}$Ca and $^{208}$Pb in units of fm$^6$.
The values are calculated by using parameters of the relativistic nuclear models, NL3\cite{nl3}
and NL-SH\cite{nlsh}, and of the non-relativistic one, SLy4\cite{sly4}.
The others show the values of the components of $H^6_{cn}$.
For the details, see the text.  
}
\label{table_hn}
\end{table}

\begin{table}
\begin{tabular}{|l|c|c||c|c|c|} \hline
$H^6_c$&
FB&
SOG&
NL3&
NL-SH&
SLy4
\\ \hline
\rule{0pt}{12pt}%
$^{40}$Ca  & $ 4.234\times 10^3[4.022]$ & $ 5.390\times 10^3[4.187]$ &$ 4.818\times 10^3$ & $4.617\times 10^3$ &$5.228\times 10^3$ \\
$^{48}$Ca  & $ 3.913\times 10^3[3.970]$ & $ 4.299\times 10^3[4.032]$ &$ 4.101\times 10^3$ & $3.971\times 10^3$ &$4.818\times 10^3$ \\ 
$^{208}$Pb & $ 5.295\times 10^4[6.128]$ & $ 5.294\times 10^4[6.128]$ &$ 5.222\times 10^4$ & $5.082\times 10^4$ &$5.401\times 10^4$ \\ \hline
\end{tabular}
\caption{
The values of the sixth moments($H^6_c$) of the charge distributions in $^{40}$Ca, $^{48}$Ca and $^{208}$Pb
in units of fm$^6$. FB and SOG indicate the Fourier-Bessel- and the sum-of-Gaussian-analyses
of the experimental data from electron scattering\cite{vries}, respectively.
In the brackets, the values of $H_c$ are listed in units of fm.
The values calculated with the MF-models are given by
$H^6_c=H^6_{cp}-H^6_{cn}$, where $H^6_{cp}$ and $H^6_{cn}$ stand for the sixth moments of
the proton and neutron charge distributions, respectively,  as listed in Table \ref{table_hp}
and \ref{table_hn}.
NL3\cite{nl3} and NL-SH\cite{nlsh} are the relativistic MF-models, while SLy4\cite{sly4}
the non-relativistic one.  For the details, see the text.  
}
 \label{table_exp}
\end{table}

\begin{table}
\begingroup
\renewcommand{\arraystretch}{1.2}
\hspace*{-1cm}%
{\setlength{\tabcolsep}{4pt}
\begin{tabular}{|c|l|l|l|l|l|l|l|} \hline
          &      &
\multicolumn{1}{c|}{$R_p$} &
\multicolumn{1}{c|}{$R_n$} &
\multicolumn{1}{c|}{$\delta R$} &
\multicolumn{1}{c|}{$Q_p$} &
\multicolumn{1}{c|}{$Q_{cp}$} & 
\multicolumn{1}{c|}{$Q_{cn}$} \\ \hline
           & RMF & $3.348(0.003)$ & $3.297(0.006)$ & $-0.050(0.009)$ & $3.635(0.015)$ & $3.785(0.014)$ & $1.512(0.004)$ \\
$^{40}$Ca  & SMF & $3.348(0.009)$ & $3.304(0.011)$ & $-0.044(0.020)$ & $3.628(0.015)$ & $3.782(0.014)$ & $1.465(0.004)$ \\ \cline{2-8}
           & Exp. & \multicolumn{3}{c|}{$R_c=3.450(0.010)$} & \multicolumn{3}{c|}{$Q_c=3.761(0.014)$} \\ \hline
           & RMF & $3.378(0.005)$ & $3.597(0.021)$  & $0.220(0.026)$ & $3.643(0.014)$ & $3.796(0.012)$ & $1.897(0.006)$ \\
$^{48}$Ca  & SMF & $3.372(0.009)$ & $3.492(0.028)$  & $0.121(0.036)$ & $3.629(0.014)$ & $3.786(0.013)$ & $1.811(0.009)$ \\ \cline{2-8}
           & Exp. & \multicolumn{3}{c|}{$R_c=3.451(0.009)$} & \multicolumn{3}{c|}{$Q_c=3.736(0.012)$} \\ \hline
           & RMF & $5.454(0.013)$ & $5.728(0.057)$ & $0.275(0.070)$ & $5.783(0.023)$ & $5.892(0.022)$ & $2.395(0.008)$ \\
$^{208}$Pb & SMF & $5.447(0.014)$ & $5.609(0.054)$ & $0.162(0.068)$ & $5.774(0.023)$ & $5.885(0.022)$ & $2.283(0.009)$ \\ \cline{2-8}
           & Exp. & \multicolumn{3}{c|}{$R_c=5.503(0.014)$} & \multicolumn{3}{c|}{$Q_c=5.851(0.022)$} \\ \hline
\end{tabular}
}
\endgroup
\caption{The results of the least squares analysis taken from Ref.\cite{kss}.
 The numbers in the parentheses for the RMF- and SMF-models denote the error
 taking account of the experimental one and the standard deviation of the calculated values
 from the least-squares line.
 All the numbers are given in units of fm. }
 \label{table_Q}
\end{table}

Before the LSA is performed in the next section,
it will be shown in this section
how each component in Eq.(\ref{cp}) and (\ref{cn}) contributes to $H^6_c$
by taking a few MF-models. 
Table \ref{table_hp} and \ref{table_hn} show the values of the components
of $H^6_{cp}$ and $H^6_{cn}$, respectively.
They are calculated using the two RMF-models, NL3\cite{nl3} and NL-SH\cite{nlsh},
and the one SMF-model, SLy4\cite{sly4}.
For the SMF models, Eq.(\ref{non}) and (\ref{nonw}) are employed in Eq.(\ref{cp})
and (\ref{cn}).
These phenomenological models are constructed so as to reproduce the fundamental quantities
of various nuclei like the binding energies, the root msr
of the charge distribution, etc.,
and have widely been used for studying other various nuclear phenomena\cite{stone,vret}.
The values of the last terms of Eq.(\ref{cp}) and (\ref{cn})
which are not listed in the tables are given, respectively, by
\begin{equation}
 \frac{35}{3}r^6_p=5.308\,,\quad 
\frac{35}{3}(r^6_+-r^6_-)\frac{N}{Z}=-1.931(^{40}{\rm Ca})\,, -2.704(^{48}{\rm Ca})\,,
-2.968(^{208}{\rm Pb})\,.
\nonumber
\end{equation}

Table \ref{table_exp} compares the MF-results with the experimental values
of $H^6_c=H^6_{cp}-H^6_{cn}$
which are evaluated by two ways. As mentioned in the previous section, the one is calculated by
using the FB-coefficients(FB), and the other
the SOG-parameters(SOG) in Ref.\cite{vries}.
Unfortunately, their experimental errors can not be estimated
from the published data\cite{vries}, where there are not enough information on 
the experimental and model-dependent errors.
Hence, the only four digits as the significant figures
are kept, so that the sixth root of $H^6_c$ in units of fm yields up to the three digits
after the decimal point, as in the brackets of Table \ref{table_exp}.
The numbers listed in Ref.\cite{vries} for the root msrs of the charge distributions
are given in the same way.
The calculated values in Table \ref{table_hp} and \ref{table_hn}, however, are shown
up to the three digits after the decimal point
so as to compare the model-predictions with one another in detail.

Let us go into more details of Table \ref{table_hp}, \ref{table_hn} and \ref{table_exp}
in order.
In Table 1 shows that the main component, $H^6_p$, contributes to $H^6_{cp}$ by more
than 75$\%$.
The second largest component is $H_{4p}$, and the sum of $H^6_p$ and $H_{4p}$ overestimates
the FB experimental values of $H^6_c$, except for the case of $^{208}$Pb by NL-SH.
All the calculated values of $H^6_{cp}$ of $^{48}$Ca and $^{208}$Pb
overestimate the FB and SOG experimental ones.
Thus, the negative contributions from the neutron charge density to $H^6_c$ are
definitely required, so as to reproduce the experimental values given
by $H^6_c=H^6_{cp}-H^6_{cn}$.
It is noted that the difference between the values of $H_{W_p}$ in the relativistic
and non-relativistic models
is not only on their absolute values, but also on their signs in $^{40}$Ca and $^{48}$Ca,
as seen in Table \ref{table_hp}.

Tables \ref{table_hp} and \ref{table_hn} show that $H^6_{cn}$ contributes to $H^6_c$ by
$9\sim 12\%$ in $^{48}$Ca, and by $4\sim 5\%$ in $^{208}$Pb.
The neutron contribution is more remarkable in $^{48}$Ca, rather than in $^{208}$Pb.
Almost a half of the value of $H_{4p}$ is cancelled by $H^6_{cn}$ in both nuclei.
The term, $H_{4n}$, accounts for more than 50$\%$ of $H^6_{cn}$. 
The contribution of $H_{2n}$ depending on $R^2_n$
is about 10$\%$ of $H^6_{cn}$ in $^{48}$Ca, and 5$\%$ in $^{208}$Pb.
In $^{208}$Pb, the value of $H^6_{cn}$ in SLy4 is smaller by more than $15\%$,
compared to that in the RMF.

Table \ref{table_exp} shows that
most of the calculated values of $H^6_c$  are between the FB- and SOG-values.
Exceptions are those of SLy4 for $^{48}$Ca and $^{208}$Pb which overestimate
the both FB- and SOG-experimental values.
The difference between those of the SLy4 and RMF models
stems  mainly from their values of $H^6_p$, as seen in Table \ref{table_hp}.
It is noticeable in Table \ref{table_exp}
that the experimental value of $H^6_c$ in $^{40}$Ca is larger than that of $^{48}$Ca.
This fact is reproduced by the three MF models in Table \ref{table_exp}.
The reason why the value of $H^6_c$ in $^{40}$Ca is larger than that
in $^{48}$Ca will be discussed in the next section.

For later discussions, the previous results obtained by the LSA
in Ref.\cite{kss}\footnote[5]{In the column of $^{48}$Ca for NL-SH in Table 2 of Ref.\cite{kss},
the value of $0.780$ for $Q_{4W_p}$ is missed. The last three numbers in the column are for
$Q_{4W_n}$, $Q^4_c$ and Exp., respectively.}are
summarized in \mbox{Table \ref{table_Q}}.
The notation, $\delta R$, stands for the neutron-skin thickness, $\delta R =R_n-R_p$,
and $Q_{cp}$ and $Q_{cn}$ represent
the fourth root of the mean fourth moments of the proton and neutron charge distributions,
respectively.
The experimental values of $R_c$ and $Q_c$ obtained by FB-method\cite{emrich}
are also listed in \mbox{Table \ref{table_Q}}.

The relationship between $Q_c$, $Q_{cp}$ and $Q_{cn}$ is given by\cite{kss},
\begin{equation}
\avr{r^4}_c =Q^4_c= Q^4_{cp}-Q^4_{cn}\,,\label{4thm}
\end{equation} 
where $Q^4_{cp}$ and $Q^4_{cn}$ are expressed as
\begin{align*}
Q^4_{cp}&= Q_{p}^4+Q_{2p}+Q_{2W_p}+Q_{4W_p}+(Q_4)_p\,, \\
Q^4_{cn}&= Q_{2n}+Q_{2W_n}+Q_{4W_n}+(Q_4)_n\,,
\end{align*}
with the notations for the protons,
\begin{equation}
Q_{2p}=\frac{10}{3}r_p^2\avr{r^2}_p,\quad
 Q_{2W_p}= \frac{10}{3}(r_p^2+\frac{3}{2M^2})\avr{r^2}_{W_p},\quad
 Q_{4W_p}= \avr{r^4}_{W_p},\quad
(Q_4)_p =\frac{5}{2}r_p^4,\nonumber
 \end{equation}
 and for the neutrons, 
 \begin{align*}
&Q_{2n}=-\frac{10}{3}(r_+^2-r_-^2) \avr{r^2}_n\frac{N}{Z},\qquad
Q_{2W_n}=-\frac{10}{3}(r_p^2+\frac{3}{2M^2}-\frac{r_+^2-r_-^2}
 {\mu_n})\avr{r^2}_{W_n}\frac{N}{Z},\qquad\\
&Q_{4W_n}=-\avr{r^4}_{W_n}\frac{N}{Z},\qquad
(Q_4)_n =-\frac{5}{2}(r_+^4-r_-^4) \frac{N}{Z}. 
 \end{align*}

Comparing Table \ref{table_Q} to Tables \ref{table_hp} and \ref{table_hn},
it is seen that the neutron contribution from $H^6_{cn}$ to $H^6_c$ is increased,
compared to that from
$Q^4_{cn}$ to $Q^4_c$, as expected. 
Ref.\cite{kss} has shown that $Q_{2n}$ is the main component of $Q^4_{cn}$,
while in $H^6_{cn}$, $H_{W_n}$
is much larger than $H_{2n}$ in $^{48}$Ca and $^{208}$Pb. The contribution
from the spin-orbit density
is more important in $H^6_{cn}$ than in $Q^4_{cn}$. Increase of the contribution
in the relativistic models, compared with that of the non-relativistic models, is due to the fact
that the spin-orbit density is enhanced by the small effective mass of nucleon in the relativistic
equation of motion\cite{ks0}.

In contrast to the values of $H^6_{cp}$ of $^{40}$Ca and $^{48}$Ca in Table \ref{table_hp}, 
those of $Q^4_{cp}$ in Table \ref{table_Q} are almost the same in the RMF- and SMF-models.
Thus, the difference between the models depends on which moment is discussed.

\section{The least-squares analysis}\label{lsa}

In the present section, the values of the various moments of the point proton and neutron
distributions allowed in the MF-framework are estimated in $^{40}$Ca, $^{48}$Ca and $^{208}$Pb,
by the LSA using the experimental values of $H^6_c$ with the SOG- and FB-methods.
The only SOG-values, however, will be indicated in all the figures below,
in order to make them simpler. The results using the FB-values will be mentioned, if necessary.

The way of the LSA is the same as that in Ref.\cite{kss} for $Q^4_c$.
First, the values of $H^6_{cp}$ and $H^6_c$ calculated in various MF models
are plotted in the ($H^6_{cp}-H^6_c$)-plane,
and the LSL between $H^6_{cp}$ and $H^6_c$ is obtained.
The value of $H^6_{cp}$ is determined by the cross point of the LSL and the line
of the experimental value of $H^6_c$. The determined value is called the LSL-value,
as mentioned in \S \ref{intro}.
Since the LSL-value of $H^6_{cp}$ is obtained with small standard deviation($\sigma$) and the
correlation coefficient($r$) nearly equal to $1$,
it will be used as the `quasi-experimental value' of $H^6_{cp}$ for other LSA
in the following steps.
Second, the values of $H^6_{p}$ and $H^6_{cp}$ calculated using the same MF models
are plotted in the ($H^6_{p}-H^6_{cp}$)-plane to obtain the LSL for them.
The LSL-value of $H^6_p$ is determined by the cross point of the LSL and the line of the
quasi-experimental value of $H^6_{cp}$ determined in the first step.
Third, the values of $Q^4_p$ are calculated, and are plotted with the calculated values
of $H^6_{cp}$ in the ($Q^4_p-H^6_{cp}$)-plane. The cross point of the LSL for ($Q^4_p-H^6_{cp}$)
and the line of the quasi-experimental value of $H^6_{cp}$ give the value of $Q^4_p$ in the
MF-framework. 
Fourth, the same procedure is repeated in order to estimate the values of $R^2_p$.
Fifth, the values of $Q^4_n$ and $H^6_{cn}$ are calculated
and are plotted in the $(Q^4_n-H^6_{cn})$-plane to obtain their
LSL. In order to obtain the quasi-experimental value of $H^6_{cn}$, 
Eq.(\ref{hc}) is employed with the quasi-experimental one of $H^6_{cp}$.
The value of $Q^4_n$ is determined by the cross point of the LSL
between $Q^4_n$ and $H^6_{cn}$ and the line of the quasi-experimental value of $H^6_{cn}$.
Finally, the values of $R^2_n$ are calculated, and are plotted with $H^6_{cn}$.
The value of $R^2_n$ in the used model-framework is determined by the cross point of the
LSL between $R^2_n$ and  $H^6_{cn}$ and the line of the quasi-experimental value of $H^6_{cn}$.

In Ref.\cite{kss} the LSL-values of $R^2_n$, $R^2_p$, and $Q^4_p$ have been estimated
for $^{40}$Ca, $^{48}$Ca and $^{208}$Pb,
by using the experimental values of $R^2_c$ and $Q^4_c$ obtained
from the FB-analysis in Refs.\cite{vries, emrich}.
In the present paper,  the LSL-values of $H^6_p$ and $Q^4_n$ will be also determined by using the
experimental values of $H^6_c$ from the both FG- and SOB-analyses.

In the following subsections, the sixth moments of the charge density in $^{40}$Ca, $^{48}$Ca
and $^{208}$Pb will be analyzed.
The MF-models employed in the present paper are the same as those in Ref.\cite{kss}.
They are 11 kinds of parameterization for nuclear interactions as RMF-models
and 9 kinds of parameterization as the SMF-models,
which are chosen arbitrarily among more than 100 models accumulated 
in nuclear physics\cite{stone,vret}. We number them, following Ref.\cite{kss} as,
1 L2\cite{sw1}, 2 NLB\cite{sw1}, 3 NLC\cite{sw1}, 4 NL1\cite{nl1}, 5 NL3\cite{nl3}, 6 NL-SH\cite{nlsh},
7 NL-Z\cite{nlz}, 8 NL-S\cite{nls}, 9 NL3II\cite{nl32}, 10 TM1\cite{tm1} and 11 FSU\cite{fsu}
for relativistic models, and
1 SKI\cite{sk1}, 2 SKII\cite{sk1},  3 SKIII\cite{sk3}, 4 SKIV\cite{sk3},
5 SkM$^\ast$\cite{skm}, 6 SLy4\cite{sly4}, 7 T6\cite{st6}, 8 SGII\cite{sg2},
and 9 Ska\cite{ska} for non-relativistic models.
The above numbering will be used throughout the present paper.

\subsection{$^{40}$Ca}

\begin{figure}[ht]
\centering{%
\includegraphics[scale=1]{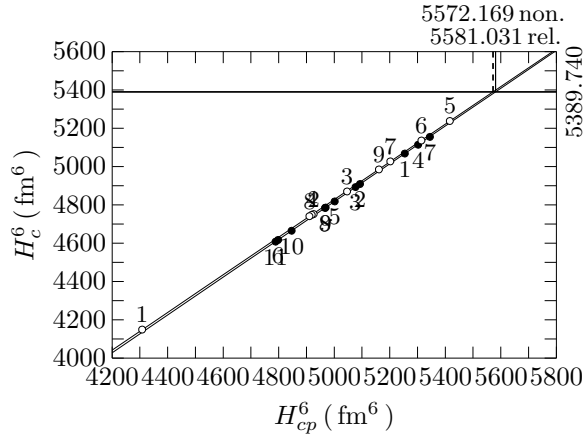}
}
\caption{The sixth moment of the charge density($H^6_c$) as a function of
the sixth moment of the proton charge density($H^6_{cp}$) in $^{40}$Ca.
The closed(open) circles are calculated in various relativistic(non-relativistic)
mean-field models. The number assigned to each circle represents a model used in
the calculations as explained in the text. The lines of the least-square fitting
are shown by the solid ones for the relativistic and non-relativistic models,
respectively. Those lines
cross the horizontal line which expresses the experimental value indicated
on the right-hand side of the figure. Their intersection points are described on the top
of the figure for the relativistic(rel.) and non-relativistic(non.) frameworks, respectively.
The values of the slope($a$), the intercept($b$) and the standard
deviation($\sigma$) of the least-squares lines are listed in Table \ref{table_lslCa40}
in the last of the present subsection. For details,
see the text.
}
\label{HcHcp-Ca40}
\end{figure}

Figure \ref{HcHcp-Ca40} shows the relationship between $H^6_{cp}$ and $H^6_c$ in $^{40}$Ca.
The calculated values with the RMF-models are indicated by the closed circles, 
while those with the SMF-ones are by the open circles.
Each circle in the figure bears the number to specify the corresponding nuclear model
mentioned in the above.
The LSLs are shown for the RMF- and SMF-models, separately, because there is no reason why
they should be taken into account together\cite{kss, toshio}.
The equations of the LSLs and their standard deviations($\sigma$)
are listed in the last of the subsection as Table \ref{table_lslCa40}.
Both RMF- and SMF-models provide
almost the same LSLs 
with the small values of $\sigma$ as $0.7718$ and $1.5701$, respectively,
and with the correlation coefficient to be about $1.0000$.
The two LSLs cross the horizontal line which expresses the SOG experimental value
of $H^6_c$ indicated
on the right-hand side of the figure. The values of the intersection points
are described on the top
of the figure for the RMF(rel.)- and SMF(non.)-frameworks, respectively.
Those are the LSL-values of $H^6_{cp} $ accepted in the RMF- and SMF-ones,
according to the LSA.
They are $5581.031$ and $5572.169$ fm$^6$
in the RMF- and SMF-frameworks, respectively,
which will be used below as a `quasi-experimental value' in $^{{40}}$Ca.
The way of the designation in Fig. \ref{HcHcp-Ca40} will be used in all the following figures
in the present paper.

It is seen in Fig. \ref{HcHcp-Ca40} that
the RMF- and SMF-models
underestimate the experimental value of $H^6_c$.
As a result, the LSL-value of $H^6_{cp}$ allowed by the MF-frameworks
is much larger than the values calculated with the MF-models used in the present LSA.
For latter discussions in \S \ref{dis}, the experimental value of $H^6_c$
obtained from the FB-analysis in Table \ref{table_exp} should be remembered.
The FB-value is $4.234\times 10^3 $ fm$^6$, which is overestimated by all the MF-models
except for SKI(1) in Fig. \ref{HcHcp-Ca40}, in contrast to the SOG-value given
by $5.390\times 10^3$ fm$^6$.

\begin{figure}[ht]
\centering{%
\includegraphics[scale=1]{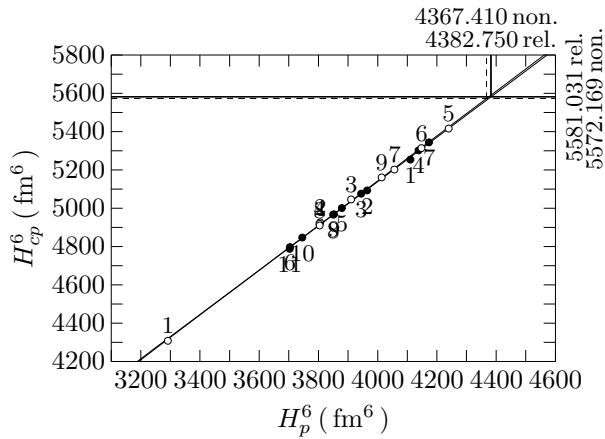}
}
\caption{The sixth moment of the proton charge density($H^6_{cp}$) as a function of
the sixth moment of the point proton distribution($H^6_p$) in $^{40}$Ca.
The solid and dashed horizontal lines indicate the quasi-experimental values of $H^6_{cp}$
for the RMF-and SKM-models, respectively. 
The designation in the figure is the same as in Fig.\ref{HcHcp-Ca40}.
For details, see the text.}
\label{HpHcp-Ca40}
\end{figure}

\begin{figure}[ht]
\centering{%
\includegraphics[scale=1]{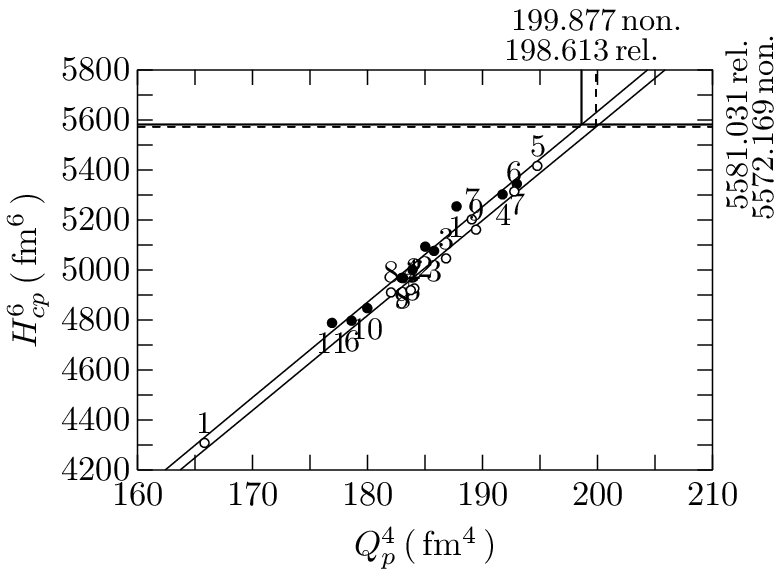}
}
\caption{The sixth moment of the proton charge density($H^6_{cp}$) as a function of
the fourth moment of the point proton distribution($Q^4_p$) in $^{40}$Ca.
The designation in the figure is the same as in Fig.\ref{HcHcp-Ca40}.
For details, see the text.
}
\label{QpHcp-Ca40}
\end{figure}

\begin{figure}[ht]
\centering{%
\includegraphics[scale=1]{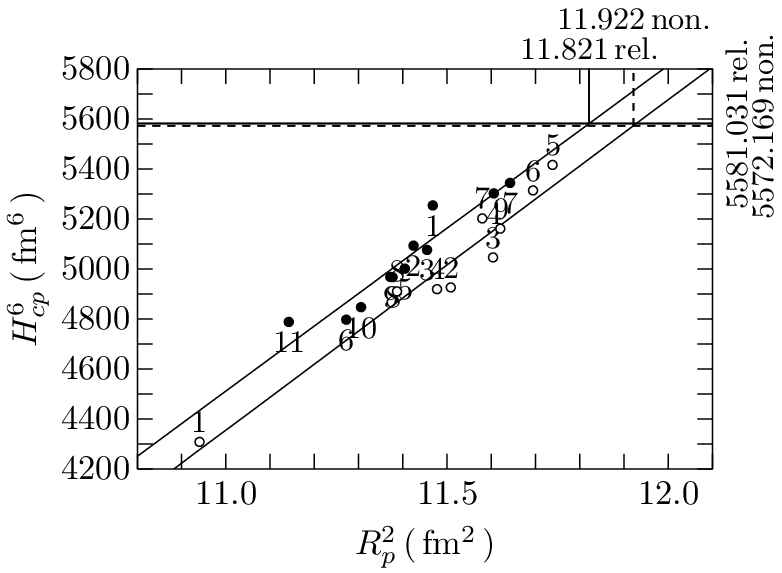}
}
\caption{The sixth moment of the proton charge density($H^6_{cp}$) as a function of
the mean square rarius of the point proton distribution($R^2_p$) in $^{40}$Ca.
The designation in the figure is the same as in Fig.\ref{HcHcp-Ca40}.
For details, see the text.
}
\label{RpHcp-Ca40}
\end{figure}

In Fig. \ref{HpHcp-Ca40} is shown $H^6_{cp}$ as a function of $H^6_p$.
Here, the LSL-values
of $H^6_{cp}$ determined in Fig. \ref{HcHcp-Ca40} are used as the quasi-experimental value,
as shown by the solid and dashed horizontal lines for the RMF- and SMF-models, respectively.
The calculated values in the RMF- and SMF-models distribute over the similar region
of the LSLs with the correlation coefficients, $r=0.9996$ and $0.9998$, respectively,
but underestimate the LSL-values of $H^6_p$.  
The order of the MF-models according to the magnitude
of $H^6_p$ is almost the same as that of Fig. \ref{HcHcp-Ca40},
and will also be seen to be the same as those of Fig. \ref{QpHcp-Ca40} for $Q^4_p$ and of
Fig. \ref{RpHcp-Ca40} for $R^2_p$.

Figure \ref{QpHcp-Ca40} shows $H^6_{cp}$ as a function of $Q^4_p$ in $^{40}$Ca.
As in Figs. \ref{HcHcp-Ca40} and \ref{HpHcp-Ca40}, both RMF- and SMF-models
underestimate the LSL-values of $Q^4_p$.
The standard deviations of the LSLs are a little large, compared with those in Figs \ref{HcHcp-Ca40}
and \ref{HpHcp-Ca40}, but their correlation coefficients are still close to $1.0000$,
as $r=0.9830$ and $0.9945$ in RMF- and SMF-frameworks, respectively.

Figure \ref{RpHcp-Ca40} shows the correlation of $R^2_p$ with $H^6_{cp}$.
As seen in Table \ref{table_hp}, $H_{2p}$ with $R^2_p$
contributes to $H^6_{cp}$ by only about $2\%$, but they are well correlated
with $r=0.9466$ and $0.9729$ in the RMF- and SMF-models, respectively.
The LSL-values of $R^2_p$ determined from $H^6_{cp}$
in Fig. \ref{RpHcp-Ca40} will be
compared to those from $R^2_c$ and $Q^4_c$ in Ref.\cite{kss} in \S \ref{dis}.

\begin{figure}[ht]
\centering{%
\includegraphics[scale=1]{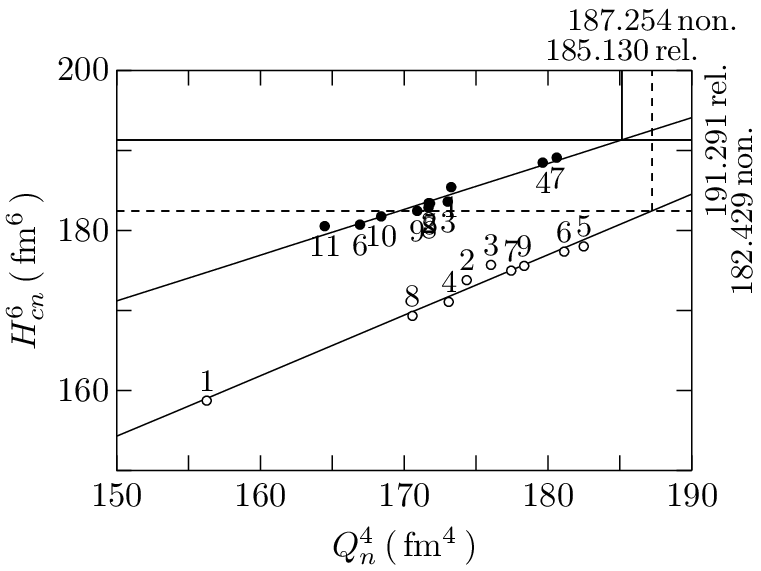}
}
\caption{The sixth moment of the neutron charge density($H^6_{cn}$) as a function of
the fourth moment of the point neutron distribution($Q^4_n$) in $^{40}$Ca.
The designation in the figure is the same as in Fig.\ref{HcHcp-Ca40}.
For details, see the text.}
\label{QnHcn-Ca40}
\end{figure}

Figure \ref{QnHcn-Ca40} shows the correlation between $Q^4_n$ and $H^6_{cn}$.
They are related as in Eq.(\ref{cn}), where $Q^4_n$ is defined as $\avr{r^4}_n$
in $H_{4n}$.
As mentioned before, the quasi-experimental values of $H^6_{cn}$ are obtained
from the experimental value of $H^6_c$ and the LSL-values of $H^6_{cp}$
by using the definition in Eq.(\ref{hc}).
The differences between the quasi-experimental values of $H^6_{cn}$ and between the LSLs
in the two frameworks in Fig. \ref{QnHcn-Ca40}
are understood in the same way as in Figs. 8 and 9 of Ref.\cite{kss}
for the ($R^2_n-Q^4_{cn}$)-correlation.
Fig. 8 in Ref.\cite{kss} shows similar differences to those in Fig. \ref{QnHcn-Ca40},
while such differences disappear in neglecting artificially the spin-orbit densities   
in the both frameworks, as exhibited in Fig. 9 in Ref.\cite{kss}.

The order of the MF-models according to the magnitude
of $H^6_{cn}$ is almost the same as that of $H^6_{cp}$ in Fig. \ref{HcHcp-Ca40}.
All the calculated values of $Q^4_n$ indicated by the circles underestimate their LSL-values
obtained as $185.130$ and $187.254$ fm$^{4}$ for the RMF- and SMF-scheme,
respectively.

\begin{figure}[ht]
\centering{%
\includegraphics[scale=1]{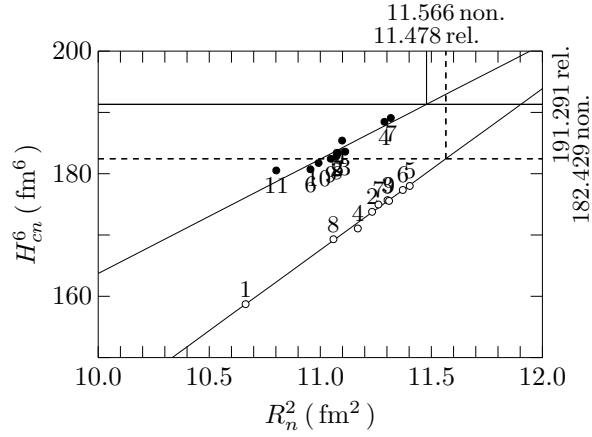}
}
\caption{The sixth moment of the neutron charge density($H^6_{cn}$) as a function of
the mean square radius of the point neutron distribution($R^2_n$) in $^{40}$Ca.
The designation in the figure is the same as in Fig.\ref{HcHcp-Ca40}.
The values of the slope($a$), the intercept($b$)
and the standard deviation($\sigma$)
of the LSL are listed in Table \ref{table_lslCa40}. For details, see the text.}
\label{RnHcn-Ca40}
\end{figure}

Figure \ref{RnHcn-Ca40} shows the correlation of $H^6_{cn}$ with $R^2_n$.
The LSL-values are underestimated by the calculated values 
in the same way as in the previous figures.
The value of the correlation coefficient of the relativistic models in this figure
is the worst among those in $^{40}$Ca, but is $r=0.9393$, while that
of the non-relativistic models $0.9978$.

The distribution of the circles in Fig. \ref{RnHcn-Ca40} is similar to
that in Fig. \ref{QnHcn-Ca40}. The difference between the LSLs is understood in the same way
as in the previous figure, as a result of the different contributions
from the spin-orbit densities in the two frameworks.
The spin-orbit densities make the difference between the two LSLs clear,
but change a little the LSL-values of $R^2_n$\cite{kss}.

Comparing Fig. \ref{RpHcp-Ca40} to Fig. \ref{RnHcn-Ca40},
the LSL-value of $R^2_p$ is larger than that of $R^2_n$
in both the RMF- and SMF-models.
In the same way, the LSL-value of $Q^4_p$ in Fig. \ref{QpHcp-Ca40} is larger
than that of $Q^4_n$ in Fig. \ref{QnHcn-Ca40}.
This result will be discussed in the following section, together with those
in $^{48}$Ca and $^{208}$Pb.

\begin{table}
 \hspace{-5mm}
\renewcommand{\arraystretch}{1.3}
\begin{tabular}{|c|c|c||r|r|r||r|r|r|} \hline
\multicolumn{3}{|c||}{$^{40}$Ca}  & \multicolumn{3}{c||}{RMF} &
\multicolumn{3}{c|}{SMF} \\ \hline
Fig.  & $y$     & $x$     & \multicolumn{1}{c|}{$a$} & \multicolumn{1}{c|}{$b$} &
\multicolumn{1}{c||}{$\sigma$} & \multicolumn{1}{c|}{$a$} & \multicolumn{1}{c|}{$b$} &
\multicolumn{1}{c|}{$\sigma$} \\ \hline
\ref{HcHcp-Ca40} &  $H^6_c$ & $H^6_{cp}$ &
$ 0.9861$ & $ -113.9240$ & $ 0.7718$ & $ 0.9823$ & $ -84.0148$ & $ 1.5701$ \\ \hline
\ref{HpHcp-Ca40} & $H^6_{cp}$ & $H^6_p$ & 
$1.1562$ & $ 513.4843$ & $ 5.2550$ & $1.1681$ & $ 470.4801$ & $ 6.3564$ \\ \hline
\ref{QpHcp-Ca40}& $H_{cp}^6$ &  $Q_p^4$ &
$ 38.1642$ & $   -1998.8749$ & $ 34.2842$ & $ 37.9642$ & $   -2016.0047$ & $ 31.9634$ \\ \hline
\ref{RpHcp-Ca40}& $H^6_{cp}$ &  $R_p^2$ &
$ 1302.8780$ & $ -9820.5972$ & $ 60.1324$ & $ 1321.1766$ & $  -10178.3905$ & $ 70.5973$ \\ \hline
\ref{QnHcn-Ca40}&  $H_{cn}^4$ &  $Q_{n}^4$ &
$0.5730$ & $ 85.2020$ & $ 0.6011$ & $0.7563$ & $ 40.8045$ & $ 0.8066$ \\ \hline
\ref{RnHcn-Ca40}&  $H_{cn}^6$ &  $R_{n}^2$ &
$ 18.6692$ & $  -22.9990$ & $ 0.9256$ & $ 26.3194$ & $ -121.9817$ & $ 0.3715$ \\ \hline
\end{tabular} 
\caption{The least-squares line $y(x)=ax+b$ and the standard deviation $\sigma$ depicted
 in Figure \ref{HcHcp-Ca40} to \ref{RnHcn-Ca40} for the relativistic(RMF)
 and the non-relativistic(SMF) models.}
 \label{table_lslCa40}
\end{table}

\subsection{$^{48}$Ca}

\begin{figure}[ht]
\centering{%
\includegraphics[scale=1]{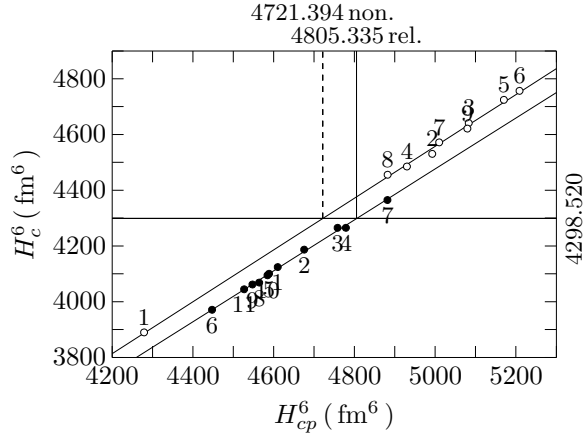}
}
\caption{The sixth moment of the charge density($H^6_c$) as a function of
the sixth moment of the proton charge density($H^6_{cp}$) in $^{48}$Ca.
The designation in the figure is the same as in Fig.\ref{HcHcp-Ca40}.
The values of the slope($a$), the intercept($b$)
and the standard deviation($\sigma$)
of the LSLs are listed in Table \ref{table_lslCa48}. For details, see the text.}
\label{HcHcp-Ca48}
\end{figure}

Figure \ref{HcHcp-Ca48} shows the relationship between $H^6_{cp}$ and $H^6_c$ in $^{48}$Ca.
The correlation coefficients are $0.9987$ and $0.9992$ for the RMF- and SMF-models, respectively,
and the equations of LSLs are listed, together with their standard deviations,
in Table \ref{table_lslCa48} at the end of this subsection.
 
The LSL-values of $H^6_{cp} $ accepted in the RMF- and SMF-models
are 4805.335 and 4721.394 fm$^6$
in the RMF- and SMF-schemes, respectively.
In contrast to the case of $^{40}$Ca, the calculated values of $H^6_c$ and $H^6_{cp}$
in the RMF-models other than NL-Z(7) underestimate the ones of their experimental
and LSL-values, while
those of the SMF-models, except for SKI(1), overestimate them.
The mean values of $H^6_c$ and $H^6_{cp}$ in the RMF models are smaller than those in the SMF
ones, whereas the LSL-value of $H^6_{cp}$ in the RMF-framework is larger than that in the
SMF-one.

\begin{figure}[ht]
\centering{%
\includegraphics[scale=1]{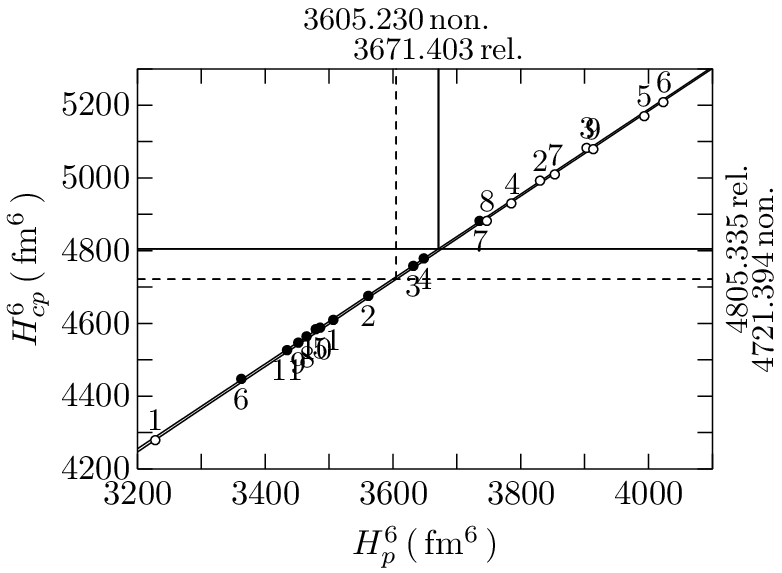}
}
\caption{The sixth moment of the proton charge density($H^6_{cp}$) as a function of
the sixth moment of the point proton distribution($H^6_p$) in $^{48}$Ca.
The designation in the figure is the same as in Fig.\ref{HcHcp-Ca40}.
For details, see the text. }
\label{HpHcp-Ca48}
\end{figure}

The experimental value, $4298.520$ fm$^6$, of $H^6_c$ in $^{48}$Ca is 
smaller than that, $5389.740$ fm$^6$, in $^{40}$Ca.  Its main reason
is due to the increasing negative contribution from
the neutron charge density.  
Moreover, it should be noticed that the LSL-values of $H^6_{cp}$ are also
smaller in $^{48}$Ca than those in $^{40}$Ca in Fig. \ref{HcHcp-Ca40}.
These results will be understood from 
Fig. \ref{HpHcp-Ca48}, which shows that
the LSL-values of $H^6_p$ is smaller in $^{48}$Ca than those in $^{40}$Ca in Fig. \ref{HpHcp-Ca40}.

In Fig. \ref{HpHcp-Ca48} is shown $H^6_{cp}$ 
as a function of $H^6_p$ in $^{48}$Ca.
The calculated values of the RMF- and SMF-models are almost on the same
line, but most of the former underestimate the LSL-value, whereas 
the latter overestimate their LSL-value.
As mentioned before, both LSL-values of $H^6_p$ are smaller
than those in $^{40}$Ca in Fig. \ref{HpHcp-Ca40}. 
The three examples in Table \ref{table_hp} shows also such a tendency, in contrast to
$H_{2p}$ with $R^2_p$.
Thus, higher moments reveal not only the contribution of the neutron charge density,
but also the detailed change of the point proton distributions in the isotopes.

\begin{figure}[ht]
\centering{%
\includegraphics[scale=1]{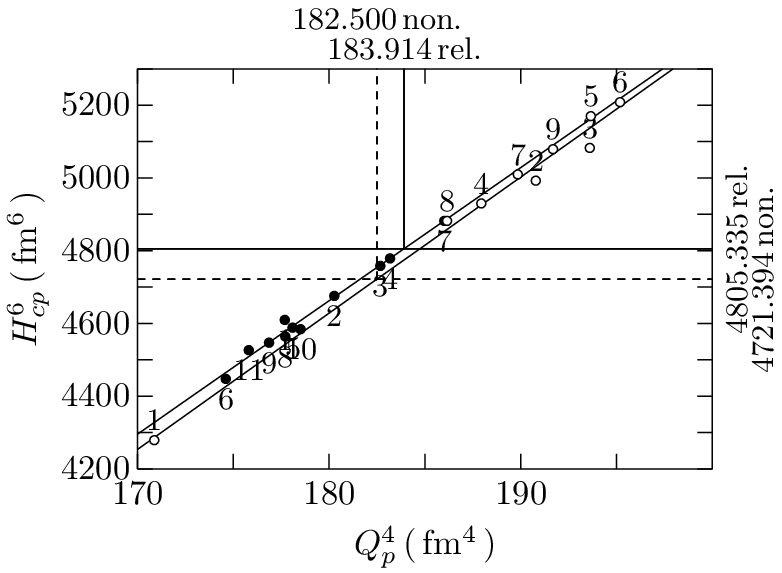}
}
\caption{The sixth moment of the proton charge density($H^6_{cp}$) as a function of
the fourth  moment of the point proton distribution($Q^4_p$) in $^{48}$Ca.
The designation in the figure is the same as in Fig.\ref{HcHcp-Ca40}.
For details, see the text. }
\label{QpHcp-Ca48}
\end{figure}

Figure \ref{QpHcp-Ca48} shows the correlation between $Q^4_p$ and $H^6_{cp}$ in $^{48}$Ca.
The RMF-models underestimate again the LSL-value, while the SMF-ones except for SKI(1)
overestimate their LSL-value in the same way as in Fig. \ref{HpHcp-Ca48}.
The LSL-values of $Q^4_p$ are given as $183.914$ and $182.500$ fm$^4$ in the RMF- and
SMF-frameworks, respectively. They are smaller than those in $^{40}$Ca
in Fig. \ref{QpHcp-Ca40} showing $198.613$ and $199.877$ fm$^4$.
It should be noted that in contrary to the SOG-value,
the FB-one in Table \ref{table_exp} yields the smaller LSL-values in $^{40}$Ca 
as $167.897$ and $168.880$ fm$^4$ in the RMF-and SMF-schemes, respectively
than $172.420$ and $171.436$ fm$^4$ in $^{48}$Ca.
As for $H^6_p$, both the SOG- and FB-values
yield the larger LSL-values in $^{40}$Ca than those in $^{48}$Ca.

\begin{figure}[ht]
\centering{%
\includegraphics[scale=1]{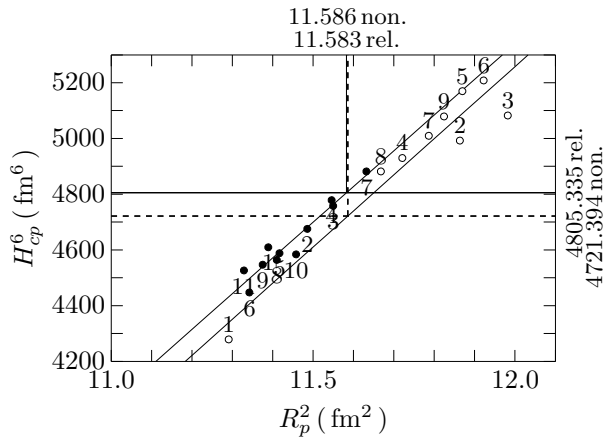}
}
\caption{The sixth moment of the proton charge density($H^6_{cp}$) as a function of
the mean square radius  of the point proton distribution($R^2_p$) in $^{48}$Ca.
The designation in the figure is the same as in Fig.\ref{HcHcp-Ca40}.
For details, see the text. }
\label{RpHcp-Ca48}
\end{figure}

Figure \ref{RpHcp-Ca48} shows $H^6_{cp}$ as a function of $R^2_p$ in $^{48}$Ca.
The contribution of $R^2_p$ through $H_{2p}$ to $H^6_{cp}$ is small,
as seen in Table \ref{table_hp},
they are well correlated with the values of $r$ to be $0.9613$ and $0.9572$ for
RMF- and SMF-models, respectively.
The distribution of the circles is almost the same as those for $H^6_{cp}$, $H^6_p$
and $Q^4_p$ in Figs. \ref{HcHcp-Ca48} to \ref{QpHcp-Ca48}.

\begin{figure}[ht]
\centering{%
\includegraphics[scale=1]{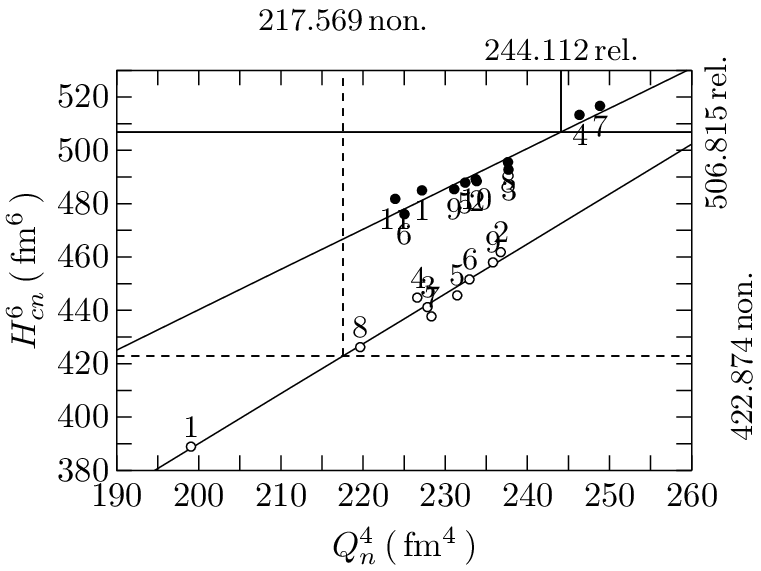}
}
\caption{The sixth moment of the neutron charge density($H^6_{cn}$) as a function of
the fourth  moment of the point neutron distribution($Q^4_n$) in $^{48}$Ca.
The designation in the figure is the same as in Fig.\ref{HcHcp-Ca40}.
 For details, see the text.}
\label{QnHcn-Ca48}
\end{figure}

The correlation of $Q^4_n$ with $H^6_{cn}$ in $^{48}$Ca is shown in Fig. \ref{QnHcn-Ca48}.
The values of $r$ are obtained as $0.9669$ and $0.9902$
for the RMF- and SMF-frameworks, respectively.
The quasi-experimental values of $H^6_{cn}$ for the two frameworks are
different, since they are obtained by Eq.(\ref{hc})
using the values of $H^6_{cp}$ given to each framework by Fig. \ref{HcHcp-Ca48}.
The difference between the two LSLs in Fig. \ref{QnHcn-Ca48} is owing mainly to the different
contributions from $H_{W_n}$ to $H^6_{cn}$ in the RMF- and SMF-models, as seen
in Table \ref{table_hn}.

On the one hand, in Fig. \ref{QpHcp-Ca48},
the calculated values of $Q^4_{p}$ in the RMF-models
are smaller than the values of the SMF-ones, except for SKI(1), 
but their LSLs yield the almost same LSL-values of $Q^4_p$. 
On the other hand, in Fig. \ref{QnHcn-Ca48}, most of the calculated values
of $Q^4_n$ are distributed over the same region between $220$ and $250$ fm$^4$,
but the LSL-values of the RMF- and SMF-frameworks are different from each other,
as shown on the top of the figure as $244.1112$ and  $217.569$ fm$^4$, respectively.
These values together with the LSL-values in Fig. \ref{QpHcp-Ca48} provide
$\delta Q=Q_n-Q_p=0.2701$ and $0.1651$ fm in the RMF- and SMF-frameworks,
respectively, which show the value of $\delta Q$ in the RMF-one is larger by about
$0.105$ fm than that in the SMF-one, in a similar way as for the neutron skin thickness
defined by $\delta R=R_n-R_p$ in Refs.\cite{kss,ks2}.

\begin{figure}[ht]
\centering{%
\includegraphics[scale=1]{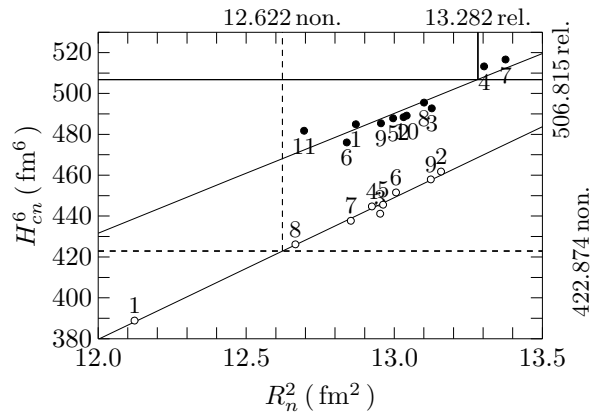}
}
\caption{The sixth moment of the neutron charge density($H^6_{cn}$) as a function of
the mean square radius of the point neutron distribution($R^2_n$) in $^{48}$Ca.
The designation in the figure is the same as in Fig.\ref{HcHcp-Ca40}.
For details, see the text.
}
\label{RnHcn-Ca48}
\end{figure}

Figure \ref{RnHcn-Ca48} shows the correlation of $R_n^2$ with $H^6_{cn}$ in $^{48}$Ca.
The values of the correlation coefficients are $0.9267$ for the RMF-models, while
$0.9958$ for the SMF-ones. The former value is the worst among those for $^{48}$Ca.
The distribution of the calculated circles is similar to that in Fig. \ref{QnHcn-Ca48}.
The line-spacing between the two LSLs is also almost the same as in Fig. \ref{QnHcn-Ca48}. 
Most of the circles by both the RMF- and SMF-models are in the same region
of $R^2_n$ around $13$ fm$^2$,
in contrast to those for the protons in Fig. \ref{RpHcp-Ca48} where all of the closed circles
distribute over the lower region of $R^2_p$ than the open circles, except for
that of SKI(1). However, whereas the LSL-values of $R^2_p$ are almost the same
in the RMF- and SMF-schemes in Fig. \ref{RpHcp-Ca48}, the LSL-values of $R^2_n$
are different from each other in Fig. \ref{RnHcn-Ca48}.
The LSL-value of $R^2_n$ in the RMF-framework is  $13.282$ fm$^2$ which is underestimated
by the most of the solid circles, while that in the SHF-one $12.622$ fm$^2$
which is overestimated by the open ones, except for SKI(1).
The two LSL-values yield the difference between $R_n$-values between the RMF-
and SMF-schemes to be about $0.092$ fm, which is almost equal to $0.105$ fm
of the difference between $\delta Q$ mentioned before.

\begin{table}
 \hspace{-5mm}
\renewcommand{\arraystretch}{1.3}
\begin{tabular}{|c|c|c||r|r|r||r|r|r|} \hline
\multicolumn{3}{|c||}{$^{48}$Ca}  & \multicolumn{3}{c||}{RMF} &
\multicolumn{3}{c|}{SMF} \\ \hline
Fig.  & $y$     & $x$     & \multicolumn{1}{c|}{$a$} & \multicolumn{1}{c|}{$b$} &
\multicolumn{1}{c||}{$\sigma$} & \multicolumn{1}{c|}{$a$} & \multicolumn{1}{c|}{$b$} &
\multicolumn{1}{c|}{$\sigma$} \\ \hline
\ref{HcHcp-Ca48} &  $H^6_c$ & $H^6_{cp}$ &
$ 0.9143$ & $ -94.8689$ & $ 5.7064$ & $ 0.9299$ & $ -91.8221$ & $ 9.4474$ \\ \hline
\ref{HpHcp-Ca48} & $H^6_{cp}$ & $H^6_p$ & 
$1.1673$ & $ 519.8788$ & $ 1.8854$ & $1.1714$ & $ 498.3485$ & $ 5.8241$ \\ \hline
\ref{QpHcp-Ca48}& $H_{cp}^6$ &  $Q_p^4$ &
$ 36.6766$ & $ -1940.0028$ & $ 14.8622$ & $ 37.4618$ & $   -2115.3859$ & $ 26.9812$ \\ \hline
\ref{RpHcp-Ca48}& $H^6_{cp}$ &  $R_p^2$ &
$ 1281.5133$ & $ -10038.2548$ & $ 33.5842$ & $ 1295.7757$ & $  -10291.3620$ & $ 75.2733$ \\ \hline
\ref{QnHcn-Ca48}&  $H_{cn}^4$ &  $Q_{n}^4$ &
$1.5108$ & $ 138.0005$ & $ 3.0357$ & $1.8709$ & $ 15.8312$ & $ 2.8757$ \\ \hline
\ref{RnHcn-Ca48}&  $H_{cn}^6$ &  $R_{n}^2$ &
$ 58.6328$ & $  -271.9630$ & $ 4.4745$ & $ 69.3482$ & $ -452.4638$ & $ 1.8884$ \\ \hline
\end{tabular} 
\caption{The least-squares line $y(x)=ax+b$ and the standard deviation $\sigma$ depicted
 in Figure \ref{HcHcp-Ca48} to \ref{RnHcn-Ca48} for the relativistic(RMF)
 and the non-relativistic(SMF) models.}
 \label{table_lslCa48}
\end{table}

\subsection{$^{208}$Pb}

\begin{figure}[ht]
\centering{%
\includegraphics[scale=1]{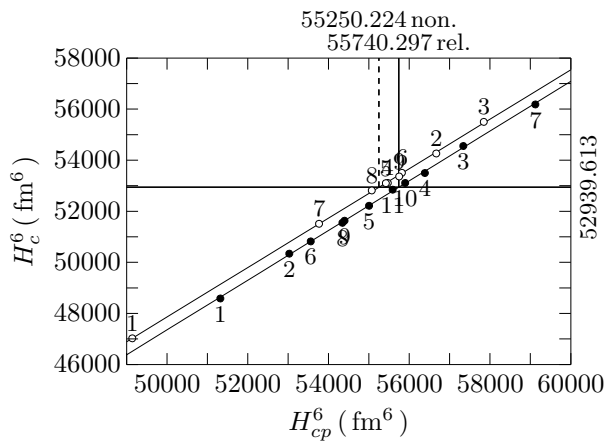}
}
\caption{The sixth moment of the charge density($H^6_c$) as a function of
the sixth moment of the point proton charge density($H^6_{cp}$) in $^{208}$Pb.
The designation in the figure is the same as in Fig.\ref{HcHcp-Ca40}.
The values of the slope($a$), the intercept($b$)
and the standard deviation($\sigma$)
of the LSLs are listed in Table \ref{table_lslPb208}. For details, see the text.
}
\label{HcHcp-Pb208}
\end{figure}

As shown in Ref.\cite{kss}, the experimental values of  $R^2_c$ and 
$Q^4_c$ of $^{208}$Pb are well reproduced by both RMF- and SMF-models,
 unlike those of $^{40}$Ca and $^{48}$Ca.
Fig. \ref{HcHcp-Pb208} shows the correlation between $H^6_{cp}$ and $H^6_c$ in $^{208}$Pb.
The circles of the calculated values are distributed over both the left-
and right-hand sides of the LSL-values of $H^6_{cp}$.
This fact implies that the mean value of $H^6_c$, $\avr{H^6_c}$,
in the MF-models well reproduce the SOG-value and that the value
of $\avr{H^6_{cp}}$ is nearly equal to its LSL-value.
The difference between the LSL-values of the RMF- and SMF-models in Fig. \ref{HcHcp-Pb208}
is about $1\%$ of their values. For $^{208}$Pb, the FB- and SOG-analyses provide almost the same
experimental value as shown in Table \ref{table_exp},
although Appendix indicates unphysical $r$-dependences in the charge distributions. 
In contrast to Fig. \ref{HcHcp-Pb208}, Figs. \ref{HcHcp-Ca40} and \ref{HcHcp-Ca48}
have shown a disagreement of the values of $\avr{H^6_c}$
in $^{40}$Ca and $^{48}$Ca with the SOG-values.

\begin{figure}[ht]
\centering{%
\includegraphics[scale=1]{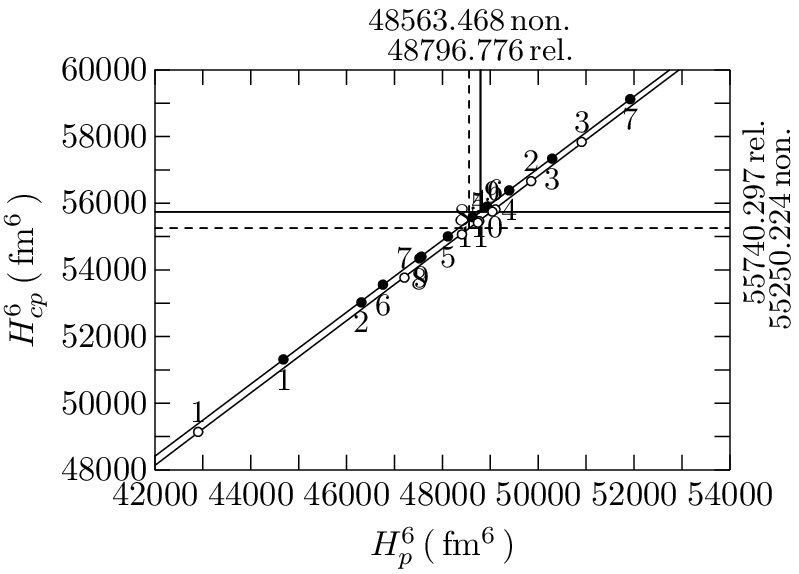}
}
\caption{The sixth moment of the proton charge density($H^6_{cp}$) as a function of
the sixth moment of the point proton distribution($H^6_p$) in $^{208}$Pb.
The designation in the figure is the same as in Fig.\ref{HcHcp-Ca40}.
For details, see the text.
}
\label{HpHcp-Pb208}
\end{figure}

The correlation of $H^6_{cp}$ with $H^6_p$ is depicted in Fig. \ref{HpHcp-Pb208},
which shows that most of the circles are near the LSL-values of $H^6_p$.
The difference between the LSL-values of $H^6_p$ in the RMF- and SMF-frameworks
is less than $1\%$ of their values, as in the previous figure .

\begin{figure}[ht]
\centering{%
\includegraphics[scale=1]{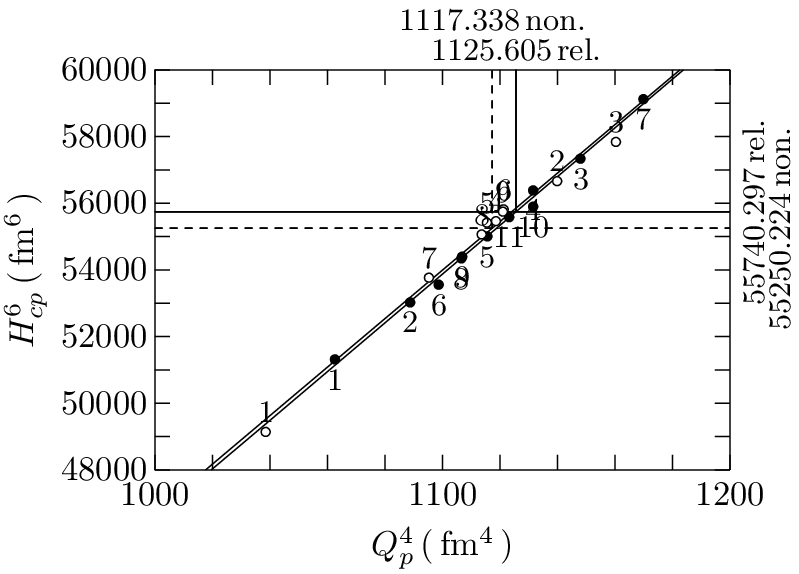}
}
\caption{The sixth moment of the proton charge density($H^6_{cp}$) as a function of
the fourth moment of the point proton distribution($Q^4_p$) in $^{208}$Pb.
The designation in the figure is the same as in Fig.\ref{HcHcp-Ca40}.
For details, see the text.
}
\label{QpHcp-Pb208}
\end{figure}

Figure \ref{QpHcp-Pb208} shows the correlation between $Q^4_p$ and $H^6_{cp}$.
The ratio of the LSL-value of $Q^4_p$ from $H^6_{cp}$ to that from
$Q^4_{cp}$ in Ref.\cite{kss} is, for example,  about 1.007 in the RMF-framework.
Such a small difference in $^{208}$Pb is owing to the fact that
the LSL-values of $Q^4_p$ are almost reproduced by the mean values calculated
with the models
in both ($Q^4_p-Q^4_{cp}$)- and ($Q^4_p-H^6_{cp}$)-correlations for $^{208}$Pb,
as will be discussed in the next section.

\begin{figure}[ht]
\centering{%
\includegraphics[scale=1]{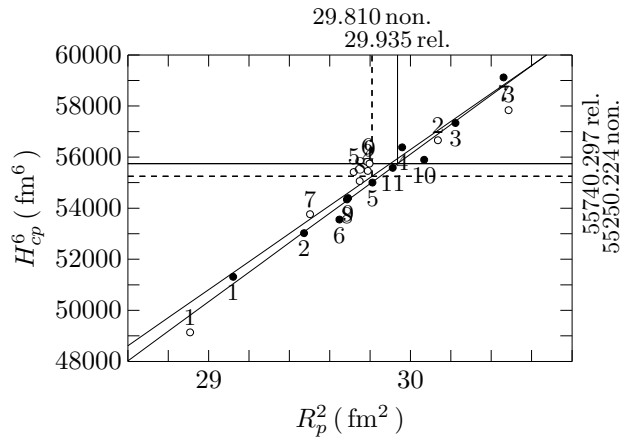}
}
\caption{The sixth moment of the proton charge density($H^6_{cp}$) as a function of
the mean square radius of the point proton distribution($R^2_p$) in $^{208}$Pb.
The designation in the figure is the same as in Fig.\ref{HcHcp-Ca40}.
For details, see the text.
}
\label{RpHcp-Pb208}
\end{figure}

In Fig. \ref{RpHcp-Pb208} is shown the correlation of $R^2_p$ with $H^6_{cp}$
in a similar way as for previous figures.
Although the contribution of $H_{2p}$ with $R^2_p$ to $H^6_{cp}$
is small as in Table \ref{table_hp}, the values of $r$ is obtained
as $0.9878$ and $0.9559$, for the RMF- and SMF-models, respectively.
The LSL-value of $R^2_p$ are $29.935$ in the RMF-framework, while $29.810$ fm$^2$
in the SMF-one. Ref.\cite{kss} provides  $29.843(0.252)$ and $29.738(0.295)$ fm$^2$
obtained from $Q^4_{cp}$, and $29.733(0.155)$ and $29.671(0.154)$ fm$^2$ from $R^2_c$
in the RMF- and SMF-frameworks, respectively. The present values are slightly larger than
the previous ones.

\begin{figure}[ht]
\centering{%
\includegraphics[scale=1]{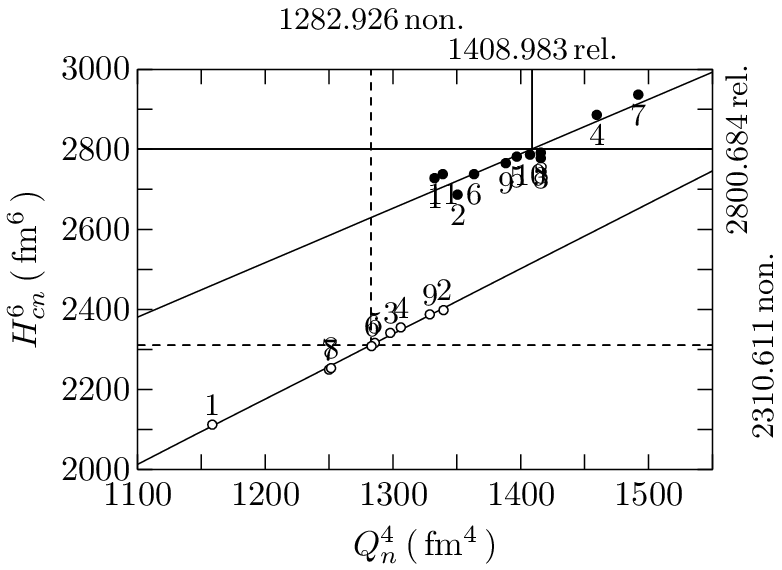}
}
\caption{The sixth moment of the neutron charge density($H^6_{cn}$)
as a function of the fourth moment of the neutron distribution($Q^4_n$)
in $^{208}$Pb.
The designation in the figure is the same as in Fig.\ref{HcHcp-Ca40}.
For details, see the text.
}
\label{QnHcn-Pb208}
\end{figure}

Figure \ref{QnHcn-Pb208} shows the correlation between $Q^4_n$ and $H^6_{cn}$.
The values of $r$ are obtained as $0.9442$ and $0.9982$ in the RMF- and SMF-models,
respectively.
The quasi-experimental value of $H^6_{cn}$ is obtained by Eq.(\ref{hc}) using
the LSL-values of $H^6_{cp}$ in Fig. \ref{HcHcp-Pb208}.
Hence, the difference between the quasi-experimental values of $H^6_{cn}$
is the same as that between the LSL-values of $H^6_{cp}$ in Fig. \ref{HcHcp-Pb208}
for the RMF- and SMF-frameworks.
In the RMF-framework, the ratio of the quasi-experimental value of $H^6_{cn}$
to the experimental one of $H^6_c$ is 0.053,
whereas that of $Q^4_{cn}$ to $Q^4_c$ is $0.028$ in Ref.\cite{kss}.
Thus, the neutron contribution is increased in $H^6_c$, compared to that in $Q^4_c$.

The spacing between the two LSLs comes from the 
difference between the values of $H_{W_n}$ in the RMF- and SMF-models,
as in Table \ref{table_hn}. 
In contrast to the case of $^{48}$Ca in Fig. \ref{QnHcn-Ca48}, the calculated values
of $Q^4_n$ in the RMF- and SMF-models are distributed over regions near their
LSL-values.

On the one hand, Fig. \ref{QpHcp-Pb208} yields almost the same LSL-values of $Q^4_p$
in the RMF- and SMF-frameworks. The difference between them is less than $1\%$.
On the other hand, those of $Q^4_n$ in Fig. \ref{QnHcn-Pb208} are different from each other
by about $9\%$.
These values provide
$\delta Q=Q_n-Q_p=0.3345$ and $0.2032$ fm in the RMF- and SMF-frameworks,
respectively, which show the value of $\delta Q$ in the RMF-one is larger by about
$0.131$ fm than that in SMF-one, in a similar way as in $^{48}$Ca.

\begin{figure}[ht]
\centering{%
\includegraphics[scale=1]{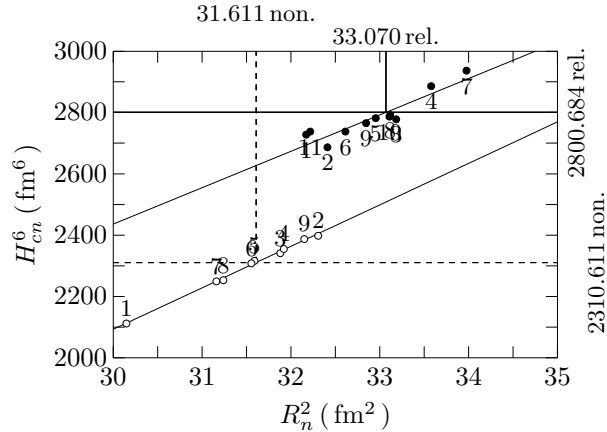}
}
\caption{The sixth moment of the neutron charge density($H^6_{cn}$)
as a function of the mean square radius of the neutron distribution($R^2_n$)
in $^{208}$Pb.
The designation in the figure is the same as in Fig.\ref{HcHcp-Ca40}.
For details, see the text.
}
\label{RnHcn-Pb208}
\end{figure}

The correlation of $R^2_n$ with $H^6_{cn}$ in $^{208}$Pb is show in Fig. \ref{RnHcn-Pb208}.
It shows that, although the main component of $H^6_{cn}$ is
$H_{4n}$ with $Q^4_n$, as seen in Table 2,
the circles for $R^2_n$ in the small component $H_{2n}$ are distributed 
in a similar way as in Fig. \ref{QnHcn-Pb208}.
The values of $r$ are $0.9252$ and $0.9980$ for the RMF- and SMF-models, respectively.
The former is the worst value among those in figures for $^{208}$Pb, while the best one
$1.0000$ for the RMF-models in Fig. \ref{HpHcp-Pb208}. 
The LSL-values of $R^2_n$ are determined to be $33.070$ and
$31.611$ fm$^2$ for the RMF- and SMF-frameworks, respectively, while they are obtained
to be $32.943(2.934)$ and $31.516(0.657)$ fm$^2$ from $Q^4_{cn}$ in Ref.\cite{kss}.
Although the value of $Q^4_{cn}$ in Ref.\cite{kss} are estimated on the basis
of the FB-analysis, the values of $R^2_n$ contain the present results within their errors.

In the previous papers\cite{kss,ks2}, it has been pointed out that whereas the LSL-values of
$R_p$ are almost the same in the RMF- and SMF-frameworks, the LSL-values of $R_n$ 
are larger by about $0.1$ fm in the RMF-framework than those in the SMF-one, in both $^{48}$Ca
and $^{208}$Pb. As a result, their $\delta R$s differ from each other by about $0.1$ fm.
In the same way, the present LSA in Fig. \ref{RnHcn-Pb208} provides
the LSL-values of $R_n$ to be larger by about $0.128$ fm in the RMF-framework than
in the SMF-one. 
Together with the LSL-values of $R_p$ in Fig. \ref{RpHcp-Pb208},
the difference between $\delta R$s
in the two frameworks is obtained as $0.117$ fm.
This $0.1$ fm difference between $\delta R$s in the RMF- and SMF-frameworks
is understood, according to Refs.\cite{toshio,ks2}.
The difference between $\delta Q$s in $^{48}$Ca and $^{208}$Pb
mentioned before may be explained in a similar way as for $\delta R$s.

\begin{table}
 \hspace{-1.2cm}
\renewcommand{\arraystretch}{1.3}
\begin{tabular}{|c|c|c||r|r|r||r|r|r|} \hline
\multicolumn{3}{|c||}{$^{208}$Pb}  & \multicolumn{3}{c||}{RMF} &
\multicolumn{3}{c|}{SMF} \\ \hline
Fig.  & $y$     & $x$     & \multicolumn{1}{c|}{$a$} & \multicolumn{1}{c|}{$b$} &
\multicolumn{1}{c||}{$\sigma$} & \multicolumn{1}{c|}{$a$} & \multicolumn{1}{c|}{$b$} &
\multicolumn{1}{c|}{$\sigma$} \\ \hline
\ref{HcHcp-Pb208} &  $H^6_c$ & $H^6_{cp}$ &
$ 0.9737$ & $ -1331.9422$ & $ 41.9577$ & $ 0.9680$ & $ -541.2208$ & $ 37.8249$ \\ \hline
\ref{HpHcp-Pb208} & $H^6_{cp}$ & $H^6_p$ & 
$1.0791$ & $ 3083.5021$ & $ 15.8481$ & $1.0841$ & $ 2603.4186$ & $ 24.8424$ \\ \hline
\ref{QpHcp-Pb208}& $H_{cp}^6$ &  $Q_p^4$ &
$ 73.0718$ & $ -26509.6358$ & $ 142.0972$ & $ 72.7496$ & $   -26035.6492$ & $ 287.1230$ \\ \hline
\ref{RpHcp-Pb208}& $H^6_{cp}$ &  $R_p^2$ &
$ 5779.9250$ & $ -117283.8075$ & $ 317.7420$ & $ 5489.9363$ & $  -108404.6446$ & $ 680.4680$ \\ \hline
\ref{QnHcn-Pb208}&  $H_{cn}^4$ &  $Q_{n}^4$ &
$1.3612$ & $ 882.7338$ & $ 22.4528$ & $1.6307$ & $ 218.5919$ & $ 5.0136$ \\ \hline
\ref{RnHcn-Pb208}&  $H_{cn}^6$ &  $R_{n}^2$ &
$ 118.7533$ & $  -1126.5053$ & $ 25.8644$ & $ 135.3781$ & $ -1968.7570$ & $ 5.2256$ \\ \hline
\end{tabular} 
\caption{The least-squares line $y(x)=ax+b$ and the standard deviation $\sigma$ depicted
 in Figure \ref{HcHcp-Pb208} to \ref{RnHcn-Pb208} for the relativistic(RMF)
 and the non-relativistic(SMF) models.}
 \label{table_lslPb208}
\end{table}

\section{Discussions}\label{dis}

\begin{table}
\begingroup
\renewcommand{\arraystretch}{1.1}
\hspace*{0mm}%
{\setlength{\tabcolsep}{3pt}
\begin{tabular}{|c|c|c|c|c|c|c|} \hline
\multicolumn{2}{|c|}{} &
\multicolumn{1}{c|}{$R^2_{c,{\rm exp}}/(\avr{R^2_c})$} &
\multicolumn{1}{c|}{$Q^4_{c,{\rm exp}}/(\avr{Q^4_c})$} &
\multicolumn{1}{c|}{$R^2_{\tau,{\rm LSL}}(a^{(2)}_{\tau,2})$} &
\multicolumn{1}{c|}{$R^2_{\tau,{\rm LSL}}(a^{(4)}_{\tau,2})$} &
\multicolumn{1}{c|}{$\avr{R^2_\tau}$} \\ \hline
          &RMF & $11.910$  & $194.734$  &$p\,11.435(1.0195)$ & $11.364(37.0126)$&$11.448$ \\ \cline{5-7}
          &    &/($11.924$) & /$(197.827)$&$n\hspace{2.5cm}$ & $12.839(16.2922)$&$13.029$ \\ \cline{2-7}
$^{48}$Ca &SMF & $11.910$  & $194.734$  &$p\,11.372(1.0000)$ & $11.336(36.7018)$&$11.772$ \\  \cline{5-7} 
          &    &/($12.310$) & /$(210.706)$&$n\hspace{2.5cm}$ & $12.147(22.1955)$&$12.866$\\ \cline{1-7}
          &RMF & $30.283$  &$1171.981$  &$p\,29.733(1.0024)$&$29.843(81.4556)$&$29.829$\\   \cline{5-7} 
          &    &/($30.379$) &/($1170.792$)&$n\hspace{2.5cm}$&$32.943(43.1093)$&$32.929$\\ \cline{2-7} 
$^{208}$Pb &SMF & $30.283$ &$1171.981$  &$p\,29.671(1.0000)$&$29.738(79.0078)$&$29.760$\\  \cline{5-7}
          &    & /($30.372$)&/($1173.747$)&$n\hspace{2.5cm}$&$31.516(46.2100)$&$31.545$\\ \cline{1-7}
\end{tabular}
}
\endgroup
\caption{The LSL-values of $R^2_\tau$ in $^{48}$Ca and $^{208}$Pb
listed as $R^2_{\tau,{\rm LSL}}(a^{(2)}_{\tau,2})$ and $R^2_{\tau,{\rm LSL}}(a^{(4)}_{\tau,2})$.
The former is obtained through the experimental values
of $R^2_{c,{\rm exp}}$, while the latter through the ones of $Q^4_{c,{\rm exp}}$.
The numbers of $a^{(2)}_{\tau,2}$ and $a^{(4)}_{\tau,2}$ in the parentheses indicate
the values of the slopes of the LSL-equations
between $R^2_\tau$ and $R^2_c$ and between $R^2_\tau$ and $Q^4_c$, respectively\cite{kss}.  
The mean values of $R^2_c$, $Q^4_c$ and $R^2_\tau$ in the relativistic
mean-field (RMF)- and non-relativistic (SMF)-models are expressed as 
$\avr{R^2_c}$, $\avr{Q^4_c}$ and $\avr{R^2_\tau}$, respectively.
The values of $R^2$ and $a^{(4)}_{\tau,2}$ are given in units of fm$^2$, 
those of $Q^4$ in units of fm$^4$, and the values of $a^{(2)}_{\tau,2}$ have no unit.
For details, see the text.}
\label{meanvalue}
\end{table}

The LSA yields for the set of the models the LSLs peculiar to its framework,
like the RMF- or the SMF-approximation.
The LSLs are expected to be inherent to the framework,
and not to depend strongly on the choice of the sample-models for the set.
Indeed, there is a following example.
The previous authors\cite{roca} has explored  
the correlation between
the slope of the asymmetry-energy density($L$)
and the neutron skin thickness($\delta R$),
by the LSA using $47$ models in the RMF- and SMF-frameworks.
They have analyzed all results in the two framework together,   
and found the LSL-equation as  $\delta R=0.147\times 10^{-2}L+0.101$ with $r=0.979$.
In the present LSA, $11$ and $9$ models are chosen arbitrarily among the RMF- and
SMF-models, respectively\cite{kss}. If these 20 models are used
in the same purpose as in Ref.\cite{roca}, one finds
the LSL-equation as $\delta R =0.151\times 10^{-2}L+0.102$ with $r=0.991$\cite{toshio},
which is almost the same result as that by 47models\cite{roca}.
Thus, it is reasonable to assume that the LSL does not depend on how the models
are chosen for the sample-set in the specified framework.

In the present paper, the moments of the charge distribution
have been analyzed. The $n$th moment($R^{(n)}_c$) is expressed in terms of the $m(\le n)$th moments
of the point proton distribution($R^{(m)}_p$) and the $m(<(n-2))$th ones of the point neutron
distribution($R^{(m)}_n$) in the nucleus\cite{ks1}. Then, the LSA is performed between $R^{(n)}_c$
and $R^{(m)}_\tau (\tau=p, n)$, whose regression(LSL)-equation is described as
\begin{equation} 
R^{(n)}_c=a^{(n)}_{\tau,m} R^{(m)}_\tau+b^{(n)}_\tau,\label{nlsl}
\end{equation}
where
$a^{(n)}_{\tau,m}$ and $b^{(n)}_{\tau,m}$ denote the slope and the intercept
of the LSL-equation, respectively.
By the definition of LSL, the above equation provides
the relationship between the mean values of the moments calculated by the models
in the set as,
\begin{equation} 
\avr{R^{(n)}_c}=a^{(n)}_{\tau,m} \avr{R^{(m)}_\tau}+b^{(n)}_\tau.\label{nmmean}
\end{equation}
When such equations for the various moments under consideration are solved simultaneously,
one obtains the average value of each calculated moment.
For example, if one is interested in the five moments, $Q^4_c$, $Q^4_{cp}$, $Q^4_{cn}$,
$R^2_p$ and $R^2_n$,
then the same number of the LSLs are selected as
\begin{align}
Q^4_c&=Q^4_{cp}-Q^4_{cn},\quad Q^4_c=\alpha_p R^2_p+\beta_p,
\quad Q^4_c=\alpha_n R^2_n+\beta_n,\\ \nonumber
Q^4_{cp}&=a_{cp}R^2_p+b_{cp}, \quad Q^4_{cn}=a_{cn}R^2_n+b_{cn}.
\end{align}
Employing the LSL-coefficients listed in Table 4 and 5 in Ref.\cite{kss},
the mean values in $^{48}$Ca and$^{208}$Pb are obtained as in Table \ref{meanvalue}.

There is a kind of the sum rule with respect to the coefficients of the LSL-equations.
Eqs.(\ref{nlsl}) and (\ref{nmmean}) provide
\begin{equation}
R^{(n)}_c-\avr{R^{(n)}_c}=a^{(n)}_{\tau,m} \left(R^{(m)}_\tau
					    -\avr{R^{(m)}_\tau}\right),\label{diffm}
\end{equation}
which is
nothing but the definition of $a^{(n)}_{\tau,m}$ in the ($R^{(m)}_\tau$, $R^{(n)}_c$)-plane.
According to the definition of the moments as in Eqs.(\ref{hc}) and (\ref{4thm}),
$R^{(n)}_c$ is assumed to be described in the form:
\begin{equation}
R^{(n)}_c=\sum_{\tau,m}\alpha^{(n)}_{\tau,m}R^{(m)}_\tau
 +\sum_{\tau,m}W^{(n,m)}_\tau + C^{(n)},\label{nm}
\end{equation}
where $\alpha^{(n)}_{\tau,m}$ stands for the coefficient coming from the nucleon-size,
$W^{(n,m)}_\tau$ the term due to the spin-orbit density, and $C^{(n)}$ term including
the nucleon-size only.
The examples of $\alpha^{(n)}_{\tau,m}$ and $W^{(n,m)}_\tau$ are given below Eqs.(\ref{cn})
and (\ref{4thm}). For example,  $\alpha^{(6)}_{p,4}=7r^2_p$ in $H_{4p}$ of $H^6_{cp}$
and $W^{(4,2)}_n=Q_{2W_n}$ in $Q^4_{cn}$ of $Q^4_c$.
In calculating Eq.(\ref{nm}) by sample-models, the mean values satisfy
\begin{equation}
\avr{R^{(n)}_c}=\sum_{\tau,m}\alpha^{(n)}_{\tau,m}\avr{R^{(m)}_\tau}
 +\sum_{\tau,m}\avr{W^{(n,m)}_\tau} + C^{(n)},\label{nmm}
\end{equation}
Eqs.(\ref{nm}) and (\ref{nmm}) provide
\begin{equation}
R^{(n)}_{c}-\avr{R^{(n)}_{c}}
=\sum_{\tau,m}\alpha^{(n)}_{\tau,m}\left(
R^{(m)}_{\tau}-\avr{R^{(m)}_\tau}\right)
+\sum_{\tau,m}\left(W^{(n,m)}_{\tau}-\avr{W^{(n,m)}_\tau}\right).\label{nmdiff}
\end{equation}
Finally, the above equation and Eq.(\ref{diffm}) yield the sum rule
on the slopes of the LSL-equations as\footnote[6]{Eq.(\ref{sumrule}) is derived, assuming
Eq.(\ref{nm}). More exact derivation is given in Appendix.}
\begin{equation}
1=\sum_{\tau,m}\frac{\alpha^{(n)}_{\tau,m}}{a^{(n)}_{\tau,m}}
 +\sum_{\tau,m}\frac{1}{a^{(n)}_{W_\tau,m}}.\label{sumrule}
\end{equation}
Here, the slope of the LSL-equation with respect to the moment of the spin-orbit
density, $a^{(n)}_{W_\tau,m}$, has been defined
in the same way as Eq.(\ref{nlsl}) for the moment, $R^{(m)}_\tau$, as
\begin{equation} 
R^{(n)}_c=a^{(n)}_{W_\tau,m} W^{(n,m)}_\tau+b^{(n)}_{W_\tau,m},\label{wlsl}
\end{equation}
which gives a similar equation to Eq.(\ref{diffm}) as
\begin{equation}
R^{(n)}_c-\avr{R^{(n)}_c}=a^{(n)}_{W_\tau,m} \left(W^{(n,m)}_\tau
					    -\avr{W^{(n,m)}_\tau}\right).\label{wdiffm}
\end{equation}
The study of the LSLs on $W^{(n,m)}_\tau$ is out of the present
purpose, but an example is found for $^{48}$Ca in Fig. 20 of Ref.\cite{kss}.
It is possible to study the LSL by expanding $W^{(n,m)}_\tau$
in terms of the moments, $\avr{r}_{W_\tau}$, of the spin-orbit density
as in Eqs.(\ref{hc}) and (\ref{4thm}).

When the moment of the charge distribution is written for the protons
and the neutrons separately as in Eqs.(\ref{hc}) and (\ref{4thm}),
\begin{equation}  
R^{(n)}_c=R^{(n)}_{cp}-R^{(n)}_{cn},
\end{equation}
and the LSL-equations are obtained as
\begin{equation}
R^{(n)}_{c\tau}=a^{(n)}_{c\tau,m} R^{(m)}_\tau+b^{(n)}_{c\tau}, \quad
R^{(n)}_{c\tau}=a^{(n)}_{cW_\tau,m} W^{(n,m)}_\tau+b^{(n)}_{cW_\tau,m},
\label{pnlsl}
\end{equation}
then the LSL-sum rule is described as
\begin{equation}
1=\sum_{m}\frac{\alpha^{(n)}_{\tau,m}}{a^{(n)}_{c\tau,m}}
 +\sum_{m}\frac{1}{a^{(n)}_{cW_\tau,m}}.\label{pnsumrule}
\end{equation}
Eq.(\ref{sumrule}) and (\ref{pnsumrule})
are convenient for examining how each moment of
the point proton and neutron distributions contributes to the $R^{(n)}_c$, and
whether or not the calculations of the LSLs are performed correctly,
as for the energy-weighted sum rule in
the random phase approximation(RPA)\cite{thouless,tsrpa,ksrpa2,ksrpa3}.

The LSL-sum rule, Eq.(\ref{sumrule}), makes clear the relationship between the experimental value
of $R^{(n)}_c$ and the LSL-values of the moments.
The LSL-value is given by Eq.(\ref{nlsl}) as
\begin{equation} 
R^{(n)}_{c,{\rm exp}}=a^{(n)}_{\tau,m} R^{(m)}_{\tau,{\rm LSL}}(R^{(n)}_{c,{\rm exp}})
+b^{(n)}_\tau, \label{lslv}
\end{equation}
where $R^{(n)}_{c,{\rm exp}}$ and $R^{(m)}_{\tau,{\rm LSL}}(R^{(n)}_{c,{\rm exp}})$
stand for the experimental value of $R^{(n)}_c$
and the LSL-value of $R^{(m)}_\tau$ determined by $R^{(n)}_{c,{\rm exp}}$, respectively.
Eqs.(\ref{nmmean}) and (\ref{lslv}) give the relationship between
the LSL-value and the mean value as
\begin{equation}
\avr{R^{(m)}_\tau}=R^{(m)}_{\tau,{\rm LSL}}(R^{(n)}_{c,{\rm exp}})+
\frac{1}{a^{(n)}_{\tau,m}}\left(\avr{R^{(n)}_c}-R^{(n)}_{c,{\rm exp}}\right).\label{diffexp}
\end{equation}
By inserting the above equation and similar equation of the moment for the spin-orbit density
into the right-hand side of Eq.(\ref{nmm}),
and using the LSL-sum rule, Eq.(\ref{sumrule}), one obtains
\begin{equation}
R^{(n)}_{c,{\rm exp}}
 =\sum_{\tau,m}\alpha^{(n)}_{\tau,m}R^{(m)}_{\tau,{\rm LSL}}(R^{(n)}_{c,{\rm exp}})
  +\sum_{\tau,m}W^{(n,m)}_{\tau,{\rm LSL}}(R^{(n)}_{c,{\rm exp}})+C^{(n)},\label{lslexp0}
\end{equation}
where the LSL-value of $W^{(n,m)}_\tau$ is defined
as $W^{(n,m)}_{\tau,{\rm LSL}}(R^{(n)}_{c,{\rm exp}})$
in the same way as in Eq.(\ref{lslv}) for $R^{(m)}_\tau$.
Thus, in spite of the fact that the $n$th moment of the charge density is
a function of several moments of the point proton and neutron distributions,
the value of each moment is provided uniquely by the LSL-value,
$R^{(m)}_{\tau,{\rm LSL}}(R^{(n)}_{c,{\rm exp}})$,
determined through the single experimental value,
$R^{(n)}_{c,{\rm exp}}$.

It should be noticed that the LSL-values in Eq.(\ref{lslexp0})
depend on $n$ of $R^{(n)}_{c,{\rm exp}}$,
as explicitly indicated in Eq.(\ref{lslv}) as $R^{(m)}_{\tau,{\rm LSL}}(R^{(n)}_{c,{\rm exp}})$.
For example. the value of $R^2_{p,{\rm LSL}}(Q^4_{c,{\rm exp}})$
is not necessary to be equal to that of $R^2_{p,{\rm LSL}}(R^2_{c,{\rm exp}})$ as
\begin{align}
R^{2}_{p,{\rm LSL}}(R^{2}_{c,{\rm exp}})&=\avr{R^{2}_p}-
\frac{1}{a^{(2)}_{p,2}}\left(\avr{R^{2}_c}-R^{2}_{c,{\rm exp}}\right),\label{diffexp2}\\
R^{2}_{p,{\rm LSL}}(Q^{4}_{c,{\rm exp}})&=\avr{R^{2}_p}-
\frac{1}{a^{(4)}_{p,2}}\left(\avr{Q^{4}_c}-Q^{4}_{c,{\rm exp}}\right).\label{diffexp4}
\end{align}
It is natural to expect that there is an ideal set $\mathcal{A}$ of the MF-models,
whose mean value  $\avr{R^2_c}_\mathcal{A}$
reproduces the experimental value of $R^2_{c,{\rm exp}}$,
because $R^2_{c,{\rm exp}}$ is usually used as one of the input-quantities to fix the free
interaction parameters.
Then, Eq.(\ref{diffexp2}) provides the LSL-value as
\begin{equation}
R^2_{p,{\rm LSL}}(R^{(2)}_{c,{\rm exp}})=\avr{R^2_p}_\mathcal{A}.\label{eq}
 \end{equation}
In order to obtain $R^2_{p,{\rm LSL}}(Q^4_{c.{\rm exp}}) =R^2_{p,{\rm LSL}}(R^2_{c.{\rm exp}})$,
Eq.(\ref{diffexp4}) implies that
the set $\mathcal{A}$ should reproduce the value of 
$Q^{4}_{c,{\rm exp}}$
also, as $\avr{Q^4_c}_\mathcal{A}=Q^{4}_{c,{\rm exp}}$.

An arbitrary set $A$ in the same framework as for $\mathcal{A}$
does not necessarily satisfy Eq.(\ref{eq}), but is expected to have the same LSL as
\begin{equation}
R^{2}_{p,{\rm LSL}}(R^{2}_{c,{\rm exp}})=\avr{R^{2}_p}_A-
\frac{1}{a^{(2)}_{p,2}}\left(\avr{R^{2}_c}_A-R^{2}_{c,{\rm exp}}\right).\label{diffexp20}
\end{equation}
In order fot the set $A$ to obtain $R^2_{p,{\rm LSL}}(Q^4_{c.{\rm exp}})
=R^2_{p,{\rm LSL}}(R^2_{c.{\rm exp}})$,  Eqs.(\ref{diffexp2}) and (\ref{diffexp4})
require
\begin{equation}
\frac{1}{a^{(4)}_{p,2}}\left(\avr{Q^{4}_c}_A-Q^{4}_{c,{\rm exp}}\right)=
 \frac{1}{a^{(2)}_{p,2}}\left(\avr{R^{2}_c}_A-R^{2}_{c,{\rm exp}}\right).\label{qr}
\end{equation}
Similar discussions can be made for other higher moments $R^{(n)}_c$ and their components
$R^{(m)}_\tau$.

The MF-framework does not assure
that the set $\mathcal{A}$ reproduces the experimental values of the
higher moments($n\ge 4$) by the mean values, and that
the set $A$ have the relationship like Eq.(\ref{qr}) 
for higher moments($n\ge 4$) also.
When the equation like Eq.(\ref{qr}) is not satisfied for $R^{(n)}_c$($n\ge 4$),
then the MF-framework with $R^2_{p,{\rm LSL}}(R^2_{c.{\rm exp}})$
fails to reproduce its experimental value,
although Eq.(\ref{lslexp0}) for $R^{(n)}_{c,{\rm exp}}$ still holds.

It is explored numerically how the LSL-values of the higher moments
are related to that of the lowest moment, $R^2_{p,{\rm LSL}}(R^2_{c,{\rm exp}})$,
used as an input-quantity.
Table \ref{meanvalue} lists the values needed for above arguments for $Q^4_c$
in $^{48}$Ca and $^{208}$Pb.
The mean values  of $R^2_c$ and $Q^4_c$  are nearly equal to the corresponding experimental
values, with one exception. The exception is in the SMF-models for $^{48}$Ca
which provide $12.310$ fm$^2$ for $\avr{R^2_c}$ against $11.910$ fm$^2$ of
the experimental value,
and $210.706$ fm$^4$ for $\avr{Q^4_c}$ against $194.734$ fm$^4$ of the experimental one. 
As a result, apart from the exception,
all the models yield each nucleus
almost the same values for 
$R^2_{p,{\rm LSL}}(R^2_{c,{\rm exp}})$, 
$R^2_{p,{\rm LSL}}(Q^4_{c,{\rm exp}})$ and $\avr{R^2_p}$, and for
$R^2_{n,{\rm LSL}}(Q^4_{c,{\rm exp}})$ and $\avr{R^2_n}$.
In these calculations, the sample-models seem to play a role of the set $\mathcal{A}$
which reproduces the experimental values of $Q^4_{c,{\rm exp}}$ also.

The results of the SMF-models for $^{48}$Ca in Table \ref{meanvalue}
seem to show approximately the relationship in Eq.(\ref{qr}).
The values of $R^2_{p,{\rm LSL}}(R^2_{c,{\rm exp}})$ and
$R^2_{p,{\rm LSL}}(Q^4_{c,{\rm exp}})$ are almost the same,
as $11.372$ and $11.336$ fm$^2$,
while the mean values of $R^2_p=11.772$ fm$^2$ are rather different from those
LSL-values. 
Table \ref{meanvalue} gives the values of the right-hand side in Eq.(\ref{qr})
to be about $0.435$ fm$^2$,
while that of the left-hand side to be about $0.400$ fm$^2$.
These values compensate the differences between $\avr{R^2_p}$
and $R^2_{p,{\rm LSL}}(Q^4_{c,{\rm exp}})$ and between  $\avr{R^2_p}$
and $R^2_{p,{\rm LSL}}(R^2_{c,{\rm exp}})$, respectively.

Once the experimental values of $H^6_c$ is determined,
then it would be possible to discuss in more detail
the $n$-dependence of the LSL-values, together with the relationship of the LSL-value,
$R^2_{\tau,{\rm LSL}}(H^6_{c,{\rm exp}})$, to the mean values.
Since there are no reliable data on $H^6_c$ yet,
the examples of the predicted values in the present LSA are listed in Table \ref{table_hc6}.
If $R^2_{p,{\rm LSL}}(H^6_{c,{\rm exp}})$ is assumed to be
equal to $R^2_{p,{\rm LSL}}(R^2_{c,{\rm exp}})$,
then the values of $H^6_{c,{\rm exp}}$ is calculated with Eq.(\ref{lslv}) and the coefficients
of the LSL-equations,
$H^6_c=aH^6_{cp}+b$, in Table \ref{table_lslCa40} to \ref{table_lslPb208}.
Comparing those values with the ones in Table \ref{table_exp},
it is seen that the predicted values are fairly different from the FB- and SOG-values.
The predicted values for $^{40}$Ca and $^{48}$Ca
are between those by FB- and SOG-method,
while the ones for $^{208}$Pb are smaller than those of FB- and SOG-analysis.
These facts may be expected from Figs. \ref{rn-Ca40}, \ref{rn-Ca48} and \ref{rn-Pb208}
in Appendix, and affect the LSL-values as in the following Table \ref{summary}.
The last column in Table \ref{table_hc6} shows the contribution of the neutrons to the
charge distributions. In $^{48}$Ca, both the RMF- and SMF-frameworks predict
the ratio of $H^6_{cn}$ to $H^6_{cp}$ to be about $10\%$,
while Ref.\cite{kss} has obtained the value of $Q^4_{cn}/Q^4_{cp}$ to be about $5\%$.

\begin{table}
 \hspace*{1.5cm}%
\begin{tabular}{|l|c|c|c|c|c|} \hline
&
&
$H^6_c$&
$H^6_{cp}$&
$H^6_{cn}$&
$H^6_{cn}/H^6_{cp}$
\\ \hline
\rule{0pt}{12pt}%
           &RMF&$4.658\times 10^3$  &$4.839\times 10^3$ &$1.812\times 10^2$ & $0.0374$ \\ \cline{2-6}
$^{40}$Ca  &SMF&$4.474\times 10^3$  &$4.640\times 10^3$ &$1.661\times 10^2$ & $0.0358$ \\ \cline{1-6}
           &RMF&$4.125\times 10^3$  &$4.616\times 10^3$ &$4.904\times 10^2$ & $0.1063$ \\ \cline{2-6} 
$^{48}$Ca  &SMF&$4.041\times 10^3$  &$4.444\times 10^3$ &$4.034\times 10^2$ & $0.0908$ \\ \cline{1-6}
           &RMF&$5.180\times 10^4$  &$5.457\times 10^4$ &$2.767\times 10^3$ & $0.0507$ \\ \cline{2-6}
$^{208}$Pb &SMF&$5.220\times 10^4$  &$5.449\times 10^4$ &$2.285\times 10^3$ & $0.0419$ \\ \hline
\end{tabular}
\caption{
The values of the sixth moments of the nuclear charge($H^6_c$), the proton charge($H^6_{cp}$), and
the neutron charge($H^6_{cn}$) distributions calculated in the relativistic(RMF) and the non-relativistic(SMF)
framework in $^{40}$Ca, $^{48}$Ca and $^{208}$Pb. They are related as $H^6_c=H^6_{cp}-H^6_{cn}$.
The mean square radii of the proton distributions in
those nuclei are determined by the least-squares method with the experimental values obtained through
Fourier-Bessel analyses of the electron scattering data\cite{kss}. The numbers are given
in units of fm$^6$.
The last row shows the ratio of $H^6_{cn}$ to $H^6_{cp}$. 
For the details, see the text.  
}
 \label{table_hc6}
\end{table}

\begin{table}
\begingroup
\renewcommand{\arraystretch}{1.2}
\hspace*{0cm}%
{\setlength{\tabcolsep}{3pt}
\begin{tabular}{|c|l|c|c|c|c|c|c|c|c|c|} \hline
 \multicolumn{2}{|c|}{}        &
\multicolumn{3}{|c|}{$^{40}$Ca}&   
\multicolumn{3}{|c|}{$^{48}$Ca}& 
\multicolumn{3}{|c|}{$^{208}$Pb} \\ \hline 
\multicolumn{2}{|c|}{}        &
\multicolumn{1}{c|}{$R_p$} &
\multicolumn{1}{c|}{$R_n$} &
\multicolumn{1}{c|}{$Q_p$} &
\multicolumn{1}{c|}{$R_p$} &
\multicolumn{1}{c|}{$R_n$} &
 \multicolumn{1}{c|}{$Q_p$} &
 \multicolumn{1}{c|}{$R_p$} &
 \multicolumn{1}{c|}{$R_n$} &
\multicolumn{1}{c|}{$Q_p$} \\ \hline 
$H^6_{c\tau}$ & RMF & $3.307$ & $3.257$ & $3.600$ & $3.355$ & $3.559$&$3.624$ &$5.471$ &$5.751$ &$5.792$   \\
            & SMF & $3.322$ & $3.283$ & $3.605$ & $3.358$ & $3.493$&$3.618$ &$5.460$ &$5.623$ &$5.782$        \\ \cline{1-11}
$Q^4_{c\tau}$ & RMF & $3.336$ & $3.268$  & $3.635$ & $3.371$ & $3.582$&$3.643$&$5.463$&$5.734$&$5.783$   \\
      &           &$(0.016)$ &(0.126)   &(0.015) & ($0.013$) & (0.112)&($0.013$)&(0.023)&(0.256)&(0.023)   \\ 
      & SMF & $3.341$ & $3.300$  & $3.628$ & $3.367$ & $3.505$ &$3.629$ &5.453&5.614&5.774  \\ 
     &      &(0.016)  &(0.032)   &(0.016) & ($0.016$) & (0.040)&$(0.014)$&(0.027)&(0.059)&(0.023)   \\ \cline{1-11}
$R^2_c$ & RMF & $3.354$ & --- & ---- & $3.382$ & ---  &----&$5.453$&----&---- \\
      &      &(0.010)  &---   &--- & ($0.009$) & ----  &----&($0.014$)&----&---- \\ 
      & SMF & $3.349$ & --- & ---- & $3.372$ & ---  &----&$5.447$&----&---- \\ 
      &      &(0.010)  &---   &--- & ($0.009$) & ---- &----&($0.014$)&----&----  \\ \cline{1-11}
\end{tabular}
}
\endgroup
\caption{The results of the least-squares analysis for the moments of
the point neutron and proton distributions in $^{40}$Ca, $^{48}$Ca and $^{208}$Pb.
The notations for the components of the moments in the second row and the first column
are defined in the text, where $\tau$ denotes the protons($\tau=p$) and neutrons($\tau=n$).
The values of $R_p$ is obtained from the analyses of $H^6_{cp}$, $Q^4_{cp}$ and $R^2_c$,
while those of $Q_p$ from $H^6_{cp}$ and $Q^4_{cp}$, and $R_n$ from $H^6_{cn}$ and $Q^4_{cn}$.
The experimental values to determine the values of $H^6_{c\tau}$, $Q^4_{c\tau}$ and $R^2_c$
are employed in the Fourier-Bessel-analyses of the electron-scattering cross sections.
The numbers in the parentheses denote the error taking account of the experimental one and 
the standard deviation of the calculated values from the least-squares line\cite{kss}.
All the numbers are given in units of fm. For details, see the text.}
 \label{summary}
\end{table}

Table \ref{summary} summarizes the LSL-values of $R_p$, $R_n$ and $Q_p$
determined by the experimental values of
$R^2_c$, $Q^4_{c\tau}$ and $H^6_{c\tau}$ in $^{40}$Ca, $^{48}$Ca and $^{208}$Pb.
Those from $R^2_c$ and $Q^4_{c\tau}$ are taken from Ref.\cite{kss}
\footnote[7]{Note that the definition of the values in Table 7 in Ref.\cite{kss}
is different from that in the present paper. Ref.\cite{kss} has determined the accepted value of the
moment by the common region between various LSL-values with errors.
Table \ref{table_Q} in the present paper follows Ref.\cite{kss}. }.
Although there are ambiguities on the experimental
values of $H^6_c$ by both FB- and SOG-analyses,
the FB-values are employed for $H^6_c$ in Table \ref{summary}, because Ref.\cite{kss} has used
the FB-ones for $R^2_c$ and $Q^4_{c\tau}$, according to the experimental values in Ref.\cite{emrich}.
Since the value of root msr $R_p$ is frequently used in the literature\cite{vries},
all the values in Table \ref{summary} are given in units of fm.
The numbers in the parentheses stand for 
the error taking account of the experimental one and 
the standard deviation of the calculated values from the least square line\cite{kss}.
It is seen that the LSL-values in $^{40}$Ca and $^{48}$Ca decrease from $R^2_c$ to $H^6_c$
in the RMF- and SMF- framework, respectively, while those in $^{208}$Pb increase inversely.
This different behavior between Ca and Pb is due to fact that the predicted values for Pb
in Table \ref{table_hc6} is smaller than the FB-values, while those for Ca larger,
as mentioned on Table \ref{table_hc6} before . 
Thus if the experimental values of $H^6_c$ are observed with small experimental errors
and those of $Q^4_c$ are determined more precisely,
understanding the MF-models would be advanced. 
The LSL-values of $H^6_p$ and $Q^4_n$ estimated from $H^6_c$ in the previous section
are not listed in Table \ref{summary}, because there are no values from $R^2_c$
and $Q^4_{c\tau}$ to be compared to.

Table \ref{summary} shows that the values of $R_p$ are larger than
the those of $R_n$ in $^{40}$Ca, yielding the negative values of the neutron-skin thickness,
$\delta R=R_n-R_p$, in contrast to the positive ones in $^{48}$Ca and $^{208}$Pb.
Figs. \ref{RpHcp-Ca40} and \ref{RnHcn-Ca40} for the SOG-values also show the same relationship
in $^{40}$Ca.
This fact may be understood qualitatively, according to the constraint
of the Hugenholtz-Van Hove(HVH) theorem
in the MF-approximation\cite{bethe,weisskopf,hvh}, as below.

Ref.\cite{toshio} has derived the following equations for $R_n$ and $R_p$
of asymmetric semi-infinite nuclear matter on the basis of the HVH theorem, 
\begin{align}
R_n&=\sqrt{\frac{3}{5}}\left(\frac{9\pi A}{8}\right)^{1/3}\frac{1}{k_{\rm F}}(1+\frac{I}{3})
 \left(1-\frac{V_c}{12J}+\frac{V_c^2}{72J^2}+\frac{L}{16K}\frac{V_c^2}{J^2}\right),
 \label{hvhn}\\
R_p&=\sqrt{\frac{3}{5}}\left(\frac{9\pi A}{8}\right)^{1/3}\frac{1}{k_{\rm F}}(1-\frac{I}{3})
 \left(1+\frac{V_c}{12J}+\frac{V_c^2}{72J^2}+\frac{L}{16K}\frac{V_c^2}{J^2}\right),\label{hvhp}
\end{align}
where, in addition to he Fermi momentum($k_{\rm F}$) and the asymmetry parameter($I=(N-Z)/A$),
$V_c(>0)$, $J$, $L$ and $K$ denote the Coulomb energy, the asymmetry-energy coefficient,
the slope of the asymmetry energy, and the incompressibility coefficient. respectively. 
All the values of the right-hand sides of the above equations are given
in each RMF- and SMF-model\cite{toshio}.
Eqs.(\ref{hvhn}) and (\ref{hvhp}) provide the neutron skin thickness
of the matter, $\delta R_M=R_n-R_p$, as
\begin{equation}
\delta R_M=\sqrt{\frac{3}{5}}\frac{2}{3}\left(\frac{9\pi A}{8}\right)^{1/3}\frac{1}{k_{\rm F}}
\left(I-\frac{1}{4}\left(\frac{V_c}{J}\right)+\frac{I}{24}\left(\frac{3L}{2K}
		  +\frac{1}{3}\right)\left(\frac{V_c}{J}\right)^2\right).\label{result}
\end{equation}
The contribution of the third term in the parenthesis to $\delta R_M$ is less than $10\%$.
The above equation is examined in detail for $^{208}$Pb in Ref.\cite{toshio},
but may be useful for rough discussions on medium-heavy nuclei also.

In $^{40}$Ca, $I=0$ implies $\delta R_M<0$, so that $R_p>R_n$,
whereas in nuclei with $N>Z$ like $^{48}$Ca and $^{208}$Pb,
we have $R_n>R_p$, because of $I>V_c/4J$\cite{toshio}.
This interpretation may also be related to the result that $Q^4_n < Q^4_p$ in $^{40}$Ca,
as mentioned before with respect to \mbox{Figs. \ref{QpHcp-Ca40} and \ref{QnHcn-Ca40}.}

The correlation between $R^2_c$ and $R^2_n$ is not required
in the definition of $R^2_c$,
but Ref.\cite{kss} has shown their well-defined LSL in the used models.
This fact implies that the correlation between $R^2_p$ and $R^2_n$ is constrained
by other requirements in the MF-models.
One of them may be Eq.(\ref{result}) which is dominated by the Coulomb- and asymmetry-energy.
Hence, when the mean values of $R^2_c$ and $Q^4_c$ are almost equal to
their experimental values, respectively, 
the LSL-value of $R^2_n$ from $R^2_c$ may be nearly equal to that from $Q^4_c$.
Indeed, for example, in $^{208}$Pb, the ($R^2_n-R^2_c$)-correlation provides the
$R^2_{n,{\rm LSL}}(R^2_{c,{\rm exp}})=32.752(0.720)$ fm$^2$,
while the ($R^2_n-Q^4_c$)-correlation
$R^2_{n,{\rm LSL}}(Q^4_{C,{\rm exp}})=32.964(0.802)$ fm$^2$ in the RMF-models\cite{kss}.
In the SMF-models,
they are $31.382(0.698)$ fm$^2$ against $31.507(716)$ fm$^2$\cite{kss}.
They are within the errors as written in the parentheses\cite{kss}.
These results shows that the values of $R^2_n$ determined by other physical quantities
like Eq.(\ref{result}) are assured to be consistent with the one in $Q^4_c$
in the MF-framework.
It is no doubt to have further information on $R^2_n$ after determining
the experimental values of $H^6_c$ in future.

\begin{table}
\begingroup
\renewcommand{\arraystretch}{1.2}
{\setlength{\tabcolsep}{4pt}
\hspace{1.7cm}
\begin{tabular}{|l|c|c|c|c|c|c|c|c|c|} \hline
 &
\multicolumn{1}{|c|}{$H^6_c({\rm Exp})$}&
 Model&
$H^6_{cp}$&
$H^6_p$&
$Q^4_p$&
$R^2_p$&
$H^6_{cn}$&
$Q^4_n$&
$R^2_n$\\ \hline 
$^{40}$Ca    & FB   & RMF & $4.409$& $3.369$ &$1.679$ &$1.092$ &$1.750$&$1.568$&$1.061$\\ \cline{3-10}
             & 4.234     & SMF & $4.395$& $3.360$ &$1.689$ &$1.103$ &$1.616$&$1.598$&$1.078$\\ \cline{2-10}
             & SOG  & RMF & $5.581$& $4.382$ &$1.999$ &$1.182$ &$1.913$&$1.851$&$1.148$\\ \cline{3-10}
             & 5.390     & SMF & $5.572$& $4.367$ &$1.986$ &$1.192$ &$1.824$&$1.873$&$1.157$\\ \hline
$^{48}$Ca    & FB   & RMF & $4.384$& $3.310$ &$1.724$ &$1.125$ &$4.707$&$2.202$&$1.267$\\ \cline{3-10}
             & 3.913     & SMF & $4.367$& $3.251$ &$1.714$ &$1.127$ &$3.938$&$2.020$&$1.220$\\ \cline{2-10}
             & SOG  & RMF & $4.805$& $3.671$ &$1.839$ &$1.158$ &$5.068$&$2.441$&$1.328$\\ \cline{3-10}
             & 4.299     & SMF & $4.721$& $3.605$ &$1.825$ &$1.159$ &$4.229$&$2.176$&$1.262$\\ \hline
$^{208}$Pb   & FB   & RMF & $55.75$& $48.81$ &$11.26$ &$2.994$ &$28.01$&$14.09$&$3.307$\\ \cline{3-10}
             & 52.95     & SMF & $55.26$& $48.57$ &$11.17$ &$2.981$ &$23.11$&$12.83$&$3.161$\\ \cline{2-10}
             & SOG  & RMF & $55.74$& $48.80$ &$11.26$ &$2.994$ &$28.01$&$14.09$&$3.307$\\ \cline{3-10} 
             & 52.94     & SMF & $55.25$& $48.56$ &$11.17$ &$2.981$ &$23.11$&$12.83$&$3.161$\\ \hline 
\end{tabular}
}
\endgroup
\caption{The LSL-values of the components of the sixth moment $H^6_c$ in $^{40}$Ca,
$^{48}$Ca and $^{208}$Pb.
The experimental values of $H^6_c$(Exp) is obtained by the two ways, which are
the Fourier-Bessel(FB)- and the sum-of-Gaussians(SOG)-analysis of the
electron-scattering cross section\cite{vries}.
The nuclear models used to estimate the LSL-values
are the relativistic (SMF) and non-relativistic(SMF) mean-field ones.
The numbers of $H^6_c$(Exp), $H^6_{cp}$ and $H^6_p$ are listed in units of $10^3\times{\rm fm}^6$,
while those of $H^6_{cn}$ in $10^2\times{\rm fm}^6$. The numbers of $Q^4_p$ and $Q^4_n$ are given
in units of $10^2\times{\rm fm}^4$, while those of $R^2_p$ and $R^2_n$ in $10\times{\rm fm}^2$.
For details, see the text.}
\label{table_fbsog}
\end{table}

As shown in Appendix, the experimental values of $H^6_c$ are not determined
by the FB- and SOG-analyses in Ref.\cite{vries}.
Ref.\cite{emrich} employed the FB-method also did not provide those values from
their experiment.
If the FB- and SOG-coefficients in Ref.\cite{vries} are used, $H^6_c$-values are calculated
as in Table \ref{table_exp}.
Table \ref{table_fbsog} lists the LSL-values of the components of $H^6_c$,
in order to show how their values are sensitive to the ways of the analyses.
The values of the components in the rows of SOG are taken from the corresponding figures
in \S \ref{lsa}. Those in the FB-rows are obtained by replacing the horizontal lines
in the figures by the FB-ones.
It is noticeable that the LSL-values of $Q^4_p$ and $R^2_p$ in $^{40}$Ca are smaller 
than those in $^{48}$Ca in the FB-analysis, while the former are larger than the latter
in the SOG-analysis. 
This result is due to the big difference between the FB- and SOG-values of $H^6_c$
in $^{40}$Ca, as $4.234\times 10^3$ and $5.390\times 10^3$ fm$^6$, respectively.

Before closing this subsection, there is a comment on
the relationship between the LAS in the present
paper and that in Refs.\cite{jlab1,jlab2,jlab3} by JLab.
They have analyzed
the cross sections for the parity-violating electron
scattering(PVES) observed at $q=0.8733$ fm$^{-1}$ in $^{48}$Ca and
at $0.476$ and $0.398$ fm$^{-1}$ in $^{208}$Pb,
$q$ being the momentum transfer from the electron to the nucleus.
On the one hand, Table \ref{table_Q} lists the values of $\delta R=R_n-R_p$
in the present method, as  $0.220(0.062)$ in the
RMF-framework and $0.121(0.036)$ fm in the SMF-one for $^{48}$Ca,
while $0.275(0.070)$ in the RMF-framework
and $0.162(0.068)$ fm in the SMF-one for $^{208}$Pb.
On the other hand, Ref.\cite{jlab3} has obtained
$\delta R= 0.121(0.050)$ for $^{48}$Ca,
and Ref.\cite{jlab2} $0.283(0.071)$ fm for $^{208}$Pb,
in the set including the both RMF- and SMF-models.
These values are all within the errors in each nucleus,
in spite of the fact that the present LSA is on the moments of the charge distribution
observed in the conventional electron scattering\cite{vries, emrich},
while JLab has analyzed the parity-violating asymmetry in the elastic scattering
of the polarized electrons from nuclei.
Moreover, the LSAs used in the present paper and Refs.\cite{jlab1,jlab2,jlab3} are different
from each other.
For understanding the difference, it is helpful to apply their method for the
conventional electron scattering, as follows.

In the present LSA, it is essential to have the reference formula, like Eqs.(\ref{hc})
and (\ref{4thm}), which shows the linear relationship between $R^{(n)}_c$
and their components. The LSAs are performed, according to those reference formulae,
and the correlations revealed by the LSLs are understood.
When the LSA similar to that of JLab is applied for the conventional electron scattering,
there is not such a formula between $R^{(n)}_c$ and the cross section.
The phase-shift calculations can not decompose
the cross section into those from $R^{(n)}_c$, much less into those from $R^{(m)}_\tau$.
Nevertheless,
the regression equation for the set of the MF-models has to be assumed for the LSA as
\begin{equation}
\sigma(q)=a^{(n)}_q R^{(n)}_c+b^{(n)}_q\label{siglsl},
\end{equation}
where $a^{(n)}_{q}$ and $b^{(n)}_q$ denote the slope and the intercept of the equation,
respectively. 
Then, Eq.(\ref{siglsl}) provides the equation in the same way as for Eq.(\ref{diffexp}),
\begin{equation}
R^{(n)}_{c,{\rm LSL}}(\sigma(q)_{\rm exp})=\avr{R^{(n)}_c}+
\frac{1}{a^{(n)}_{q}}\left(\sigma(q)_{\rm exp}-\avr{\sigma(q)}
			   \right).\label{diffexpc} 
\end{equation}
Note that Eq.(\ref{diffexpc}) does not imply
\begin{equation}
R^{(n)}_{c,{\rm LSL}}(\sigma(q)_{\rm exp})
=R^{(n)}_{c,{\rm exp}},\label{ass1}
\end{equation}
but in assuming the above equality,
Eqs.(\ref{diffexp}) and (\ref{diffexpc}) provide
\begin{equation}
\left(\sigma(q)_{\rm exp}-\avr{\sigma(q)}\right)
=a^{(n)}_q a^{(n)}_{\tau,m}\left(R^{(m)}_{\tau,{\rm LSL}}(R^{(n)}_{c,{\rm exp}})
-\avr{R^{(m)}_\tau}\right).\label{ass2}
\end{equation}
It is, however, unreasonable that all the value of
$R^{(m)}_{\tau,{\rm LSL}}(R^{(n)}_{c,{\rm exp}})$
is determined for all $n$ by a single experimental value of the cross section
at any value of $q$.
This result comes from the assumption in Eq.(\ref{siglsl}) with no reference formula.
Thus, there seems not to be a obvious relationship between the present LSA and that of JLab. 

The similarity between the values of $\delta R$s in JLab and 
the present analysis may indicate that Eqs.(\ref{siglsl}) and (\ref{ass1})
hold for the small values of $n$ at low $q$,
although the plane-wave Born approximation to the estimation of the cross section
is not appropriate for the present discussions\cite{deforest,jlab1}.
If Eq.(\ref{ass1}) is valid for $n=2$ and $4$,
the LSA between $R^{(m)}_\tau (m=2)$ and $\sigma(q)$
would provide the value of $\delta R$.
One of the ways to verify Eq.(\ref{ass1}) for small values of $n$
in the conventional electron scattering 
is to explore if the right-hand side of Eq.(\ref{diffexpc}) is $q$-independent,
in addition to the fact that
the value of $R^{(n)}_{c,{\rm LSL}}(\sigma(q)_{\rm exp})$ 
is almost equal to the one of $R^{(n)}_{c,{\rm exp}}$ which is known already
in other experiments\cite{vries,emrich}.
Eq.(\ref{ass1}) is also realized,
when the value of $R^{(n)}_{c,{\rm exp}}$ 
is reproduced by the model-set as
\begin{equation}
R^{(n)}_{c,{\rm exp}}=\avr{R^{(n)}_c}. \label{rcrc}
\end{equation}
Here the mean value of $R^{(n)}_c$ is calculated by the charge density:
\begin{equation}
\rho_c(r)=\frac{1}{N}\sum_{i=1}^N \rho_{ci}(r), \label{mden0}
\end{equation}
where $\rho_{ci}(r)$ denotes the charge density calculated with the model $i$ in the set
of the $N$ models.
If $\rho_c(r)$ reproduces all the values of $R^{(n)}_{c,{\rm exp}}$s under consideration,
the phase-shift calculation with the use of Eq.(\ref{mden0}) is expected to explain
the experimental value of the cross section as
\begin{equation}
 \sigma(q)_{\rm exp}=\avr{\sigma(q)}.\label{cccc}
\end{equation}
Then, Eq.(\ref{rcrc}) and (\ref{cccc}) in Eq.(\ref{diffexpc}) yield Eq.(\ref{ass1}).

Once Eq.(\ref{ass1}) is found to be valid for $n=2$,
it would be examined  for $Q^4_c$,
and also the related $R^{(m)}_\tau$ of the point proton and neutron distributions.
It is interesting to extend the method to $H^6_c$ also,
for which there is no way to extract the experimental values at present.
New accurate experimental data
at low $q$-region $\le 0.8$ fm$^{-1}$ for $^{48}$Ca
and $\le 0.5$ fm$^{-1}$ for $^{208}$Pb\cite{ks1} are desired for
these considerations,
in addition to the previous ones with higher $q$\cite{vries,emrich,frosch}. 

The conventional electron-scattering experiment may be performed much more easily
than the time-consuming PVES-experiment.
Moreover, in order to see the validity of Eq.(\ref{diffexpc}) only,
it is not necessary to determine the absolute values
of the cross section which require usually other efforts.
Suppose that the  cross section observed by experiment,
$\sigma(q)_{\rm exp}$, is proportional to
the cross section, $\sigma(q)_{\rm ab}$, whose absolute value is assumed to
be known, as
\begin{equation}
\sigma(q)_{\rm exp}=c\sigma(q)_{\rm ab}, \quad (c={\rm constant}).
\end{equation}
Then,
if $R^{(n)}_{c,{\rm LSL}}(\sigma(q)_{\rm ab})=R^{(n)}_{c,{\rm LSL}}(\sigma(q_0)_{\rm ab})$,
Eq.(\ref{diffexpc}) is rewritten in terms of the ratio of the cross sections,
$A(q)=\sigma(q)_{\rm exp}/\sigma(q_0)_{\rm exp}$, as
\begin{equation}
R^{(n)}_{c,{\rm LSL}}(\sigma(q)_{\rm ab})=\avr{R^{(n)}_c}+
\frac{\avr{\sigma(q_0)}A(q)-\avr{\sigma(q)}}
{a_q-a_{q_0}A(q)}, \quad (q_0\ne q),\label{sigma1}
\end{equation}
where $q_0$ stands for one of the momentum transfer which is arbitrarily chosen,
avoiding the region $q_0\approx q$. 
When the value of $R^{(n)}_{c,{\rm LSL}}(\sigma(q)_{\rm ab})$ is determined,
the constant, $c$, is given by
\begin{equation}
c=\frac{\avr{\sigma(q_0)}-\avr{\sigma(q)}+(a_{q_0}-a_q)
(R^{(n)}_{c,{\rm LSL}}(\sigma(q)_{\rm exp})-\avr{R^{(n)}_c})}
{\avr{\sigma(q_0)}-\avr{\sigma(q)}+(a_{q_0}-a_q)
(R^{(n)}_{c,{\rm LSL}}(\sigma(q)_{\rm ab})-\avr{R^{(n)}_c})}.\label{sigma2}
\end{equation}
If $R^{(n)}_{c,{\rm LSL}}(\sigma(q)_{\rm ab})=\avr{R^{(n)}_c}$, then
 Eqs.(\ref{sigma1}) and (\ref{sigma2}) provide, respectively,
 \begin{equation}
A(q)=\frac{\avr{\sigma(q)}}{\avr{\sigma(q_0)}}=\frac{\sigma(q)_{\rm exp}}
{\sigma(q_0)_{\rm exp}}, \quad
c=1+\frac{(a_{q_0}-a_q)
(R^{(n)}_{c,{\rm LSL}}(\sigma(q)_{\rm exp})-\avr{R^{(n)}_c})}
{\avr{\sigma(q_0)}-\avr{\sigma(q)}}.
\end{equation}
Eq.(\ref{sigma2}) gives also $c=1$
for $R^{(n)}_{c,{\rm LSL}}(\sigma(q)_{\rm exp})
=R^{(n)}_{c,{\rm LSL}}(\sigma(q)_{\rm ab})$, as it should be.

It is noted that 
if $N\rightarrow \infty$ in Eq.(\ref{mden0}),
then it would provide the `model-independent' charge density,
as far as the moments of the charge density is concerned,
but in the model-framework used. The real model-independent
densities of the point proton and neutron distributions can not be obtained,
because the definition of the moment of the charge distributions in RMF
and SMF-frameworks
are different from each other.
Model-dependent contributions to the charge density from the relativistic effects
including spin-orbit densities are different in the two frameworks. As a result,
for example,
there may remain the $0.1$ fm difference
between $R^2_n$s of the RMF- and SMF-framework in the LSA\cite{kss,ks2}.

\section{Summary}\label{sum}

The $n$th moment of the charge distribution, $R^{(n)}_c$, depends not only on
the $m(\le n)$th moment of the point proton distribution($R^{(m)}_p$),
but also on the $m(\le (n-2))$th moment of the point neutron one($R^{(m)}_n$)\cite{ks1}.
The experimental value of $R^{(n)}_c$ is determined through electron scattering where
the electromagnetic interaction and reaction mechanism
are well understood\cite{bd,deforest}.
It is noticeable that the point neutron distribution is
investigated, comparing with same experiment as that for  the point proton distribution.
Both the point proton and the point neutron distribution are most fundamental quantities
for description of nuclei.

In order to decompose the experimental value of $R^{(n)}_c$
into the components, $R^{(m)}_\tau (\tau=p, n)$,
it is necessary to use nuclear models.
At present, it is not avoidable for most of the investigations on nuclear structure
to rely on phenomenological models more or less,
and it seems to be difficult to select one definitive model among them.
In this paper, the mean-field(MF) models have been utilized.
There are three reasons why the MF-models are employed.
First,
they have played an important role in understanding fundamental properties of nuclei,
since the early period of nuclear physics\cite{bm}.
Second,
for at least the last 50 years,
it is continued to improve the MF-models in the same framework,
aiming to explain various static and dynamical aspects of nuclei\cite{nl3,sly4}.
The one is the non-relativistic framework(SMF)
based on the work of Vautherin and Brink\cite{sk1},
and the other is the relativistic one(RMF) by Horowitz and Serot\cite{hs}.
As a result, many models with different phenomenological interactions
have been accumulated in the same MF-framework\cite{stone, nl3}.
In the case of the SMF-framework, there are more than $100$ types of the models\cite{stone}. 
Third,
taking advantage of the abundant accumulation
of the elaborate models,
the least-squares method(LSA) can be used for estimating each value
of $R^{(m)}_\tau$  in the model-framework from the single experimental value, $R^{(n)}_c$,
as shown in Eq.(\ref{lslexp0}).
The LSA reveals the constraint inherent in the model-framework through the
least-squares line(LSL).
Although the contributions from the neutrons to $R^{(n)}_c$ are about $5\%$ in $Q^4_c$
and $10\%$ in $H^6_c$, they are not a small amount of the correction
for the present sophisticated MF-models and appear clearly in the LSLs\cite{kss}.

The estimated results of $R^{(m)}_\tau$ are summarized in Table \ref{summary}, where
the values of $R^{(m)}_\tau$ related to the sixth moment($H^6_c$) of the charge
distribution obtained in the present paper are listed,
in addition to those from the second($R^2_c$) and the fourth moment($Q^4_c$) in Ref.\cite{kss}.
The values from $H^6_c$, however, are shown for reference,
because their experimental values has not been determined by the Fourier-Bessel-
and sum-of-Gaussians-coefficients listed in Ref.\cite{vries}, as discussed in Appendix.
Since higher moments receive more contributions from the neutrons, new methods to determine
their experimental values are desired for investigating the moments of
the point neutron distribution in detail.    
In addition, information on higher moments would improve our understanding structure
of the nuclear surface.

Finally, one comment is added  on the MF-models.
It may be continued to improve the MF-models from now on,
because there is no basic principle to determine the interaction parameters.
For example, no-sea-approximation in the RMF-models neglect the divergence problem
due to nucleon-antinucleon states\cite{fur,ma}.
Those excitations strongly affect the value of the incompressibility($K$)\cite{ks50}
which is frequently used as an input-quantity in the RMF-models\cite{nl3}.
The neglect of the divergence terms changes
the attractive interaction due to the $\sigma$-exchange to
the repulsive one unphysically\cite{ksrpa3}.
Moreover, the neglect leads to the violation of the energy-weighted sum
rule in the RPA\cite{ksrpa2}.
A proper treatment of the divergence may require additional interaction-parameters
in the RMF-models in future.
In the non-relativistic models,
the effective mass is determined by the zero-range interactions with the momentum-dependence
\cite{sly4}. Their parameters are related to
the enhancement factor $\kappa$ of the sum of the dipole excitation strengths\cite{sly42}.
The value of $\kappa$ observed in photo-excitations are mainly
due to charge-exchange interactions with finite range\cite{bm,tsrpa}.
It is not clear, however, how many $\%$ of the value of $\kappa$ should be explained
by the momentum-dependent zero-range forces\cite{bm,tsrpa}.
Moreover, the momentum-dependent zero-range forces may include the finite range
effects without charge exchange operators\cite{tsrpa}.
Thus, there seems to have possibility to require additional parameters
in both the RMF- and SMF-models for improvement.
In constructing a new MF-model, 
the LSL and LSL-values obtained in this paper
may offer a guide for reproducing the moments, where
it is enough to take care of the LSL-values of $R^{(m)}_\tau$s
of the point neutron and proton distributions
in each framework,
instead of the experimental values of $R^{(n)}_c$s with complicated corrections.

\vspace{3mm}
\noindent
{\bf Acknowledgments}

The author would like to thank Professor H. Kurasawa and Professor T. Suda
for useful discussions. Most of the numerical calculations are indebted to them.
This work was supported by JSPS KAKENHI Grant Numbers JP22K18706.

\vspace{5mm}
\noindent
\large{\bf{Aappendix}}

\vspace{2mm}
\noindent
\large{\bf A. {Sum rule}}

\vspace{3mm}

\renewcommand{\theequation}{A.\arabic{equation}}
\setcounter{equation}{0}

Eq.(\ref{sumrule}) is derived, assuming Eq.(\ref{nm}). More exact derivation of the sum rule
is provided as follows. For the LSL-equation:
\begin{equation}
R^{(n)}_c=a^{(n)}_{\tau,m}R^{(m)}_\tau+b^{(n)}_{\tau,m},
\end{equation}
the element $(R^{(m)}_{\tau,i}, R^{(n)}_{c,i})$ by the model $i$ satisfies
\begin{equation} 
R^{(n)}_{c,i}=a^{(n)}_{\tau,m}R^{(m)}_{\tau,i}+b^{(n)}_{\tau,m}+\delta^{(m,n)}_{\tau,i},
 \end{equation}
and by the definition of LSL, their mean values is given by Eq.(\ref{nmmean}) as
\begin{equation}
\avr{R^{(n)}_c}=a^{(n)}_{\tau,m}\avr{R^{(m)}_\tau}+b^{(n)}_{\tau,m},
\end{equation}
where $\delta^{(m,n)}_{\tau,i}$ stand for the deviation from the LSL.
The above two equations yield
\begin{equation}
R^{(m)}_{\tau,i}-\avr{R^{(m)}_\tau}=\frac{R^{(n)}_{c,i}-\avr{R^{(n)}_c}}{a^{(n)}_{\tau,m}}
\left(1-\frac{\delta^{(m,n)}_{\tau,i}}{R^{(n)}_{c,i}-\avr{R^{(n)}_c}}\right).\label{diffri}
\end{equation}
In the same way, one has for the spin-orbit density
\begin{equation}
W^{(n,m)}_{\tau,i}-\avr{W^{(n,m)}_\tau}=\frac{R^{(n)}_{c,i}-\avr{R^{(n)}_c}}{a^{(n)}_{W_\tau,m}}
\left(1-\frac{\delta^{(m,n)}_{W_\tau,i}}{R^{(n)}_{c,i}-\avr{R^{(n)}_c}}\right)\label{diffwi}
\end{equation}
with the deviation from the LSL by $\delta^{(m,n)}_{W_\tau,i}$.

Now, according to the definition, $R^{(n)}_{c,i}$ is provided by
\begin{equation}
R^{(n)}_{c,i}=\sum_{\tau,m}\alpha^{(n)}_{\tau,m}R^{(m)}_{\tau,i}
 +\sum_{\tau,m}W^{(n,m)}_{\tau,i} + C^{(n)},\label{nmi}
\end{equation}
and Eq.(\ref{nmm}) holds as
\begin{equation}
\avr{R^{(n)}_c}=\sum_{\tau,m}\alpha^{(n)}_{\tau,m}\avr{R^{(m)}_\tau}
 +\sum_{\tau,m}\avr{W^{(n,m)}_\tau} + C^{(n)}.\label{nmm0}
\end{equation}
Then, the above two equations give
\begin{equation}
R^{(n)}_{c,i}-\avr{R^{(n)}_{c}}
=\sum_{\tau,m}\alpha^{(n)}_{\tau,m}\left(
R^{(m)}_{\tau,i}-\avr{R^{(m)}_\tau}\right)
+\sum_{\tau,m}\left(W^{(n,m)}_{\tau,i}-\avr{W^{(n,m)}_{\tau,i}}\right).\label{nmdiff0}
\end{equation}
Inserting Eqs.(\ref{diffri}) and(\ref{diffwi}) into Eq.(\ref{nmdiff0}),
finally, the sum rule on the slopes of the LSL-equations is written as
\begin{equation}
1=\sum_{\tau,m}\frac{\alpha^{(n)}_{\tau,m}}{a^{(n)}_{\tau,m}}
\left(1-\frac{\delta^{(m,n)}_{\tau,i}}{R^{(n)}_{c,i}-\avr{R^{(n)}_c}}\right) 
 +\sum_{\tau,m}\frac{1}{a^{(n)}_{W_\tau,m}}
 \left(1-\frac{\delta^{(m,n)}_{W_\tau,i}}{R^{(n)}_{c,i}-\avr{R^{(n)}_c}}\right).\label{sumrule0}
\end{equation}
Thus, Eq.(\ref{sumrule}) is obtained , when $\delta^{(m,n)}_{\tau,i}, \delta^{(m,n)}_{W_\tau,i},
\ll(R^{(n)}_{c,i}-\avr{R^{(n)}_c})$.
Eq.(\ref{sumrule0}) implies that Eq.(\ref{lslexp0}) is also affected by  
$\delta^{(m,n)}_{\tau,i}$ and $\delta^{(m,n)}_{W_\tau,i}$.

\vspace{5mm}
\noindent
\large{\bf B. {Fourier-Bessel and Sum-of-Gaussians analyses}}

\vspace{3mm}

\renewcommand{\theequation}{A.\arabic{equation}}
\setcounter{equation}{0}

\begin{table}
 \hspace*{-0.2cm}%
\begin{tabular}{|c|c|c|c|c||c|c|c|c|c|} \hline
\rule{0pt}{12pt}
$^{40}$Ca  &
$R^2_c$&
$Q^4_c$&
$H^6_c$&
$\avr{r^8}_c$&
 $^{48}$Ca & $R^2_c$   &$Q^4_c$ & $H^6_c$  &  $\avr{r^8}_c$  \\ \hline
\rule{0pt}{12pt}%
NL3        & $12.06$ & 209.9 & $4818$ & $142700$  &NL3 &$11.89$ &$196.7$ &$4101$ & $103600$         \\
           &($3.473$) &($3.806$) & ($4.110$) &  ($4.409$) & &($3.448$) &($3.745$) &($4.001$) &(4.236$$)  \\ \hline
FB         & $11.90$ & $ 200.0$  &$ 4234$ & $ 106900$  &FB &$11.91$ &$194.7$ &$3913$ &$91690$   \\   
           & ($3.450$) & ($3.761$) & ($4.022$)  &(4.252$$) & &($3.451$) &($3.735$) &($3.970$) & ($4.171$)  \\
SOG        & $ 12.11$ & $ 216.5$ & $ 5390$ & $189300$  &SOG &$11.97$ &$200.5$ &$4299$ &$116100$  \\ 
           &  ($3.480$) & ($3.836$) &($4.187$) & ($4.567$) & &($3.460$) &($3.763$) &($4.032$) & ($4.296$)  \\ \hline
Ratio&0.9914&0.9804&0.9606&0.9310&Ratio &0.9974&0.9926&0.9846&0.9709\\ \hline
\end{tabular}
\caption{
The values of the moments of the charge densities in $^{40}$Ca and $^{48}$Ca calculated by
the relativistic mean-field model(NL3\cite{nl3}) and by the Fourier-Bessel(FB)- and sum-of-Gaussians(SOG)-analyses of electron-scattering data\cite{vries}.
The numbers of $R_c$, $Q_c$, $H_c$ and the eighth root of $\avr{r^8}_c$ are given in units of fm
in the parentheses. The last row shows the ratio of the FB-number to the SOG-one in the parentheses.
For the notations and details, see the text.   
}
\label{table_nfs}
\end{table}

\begin{figure}[ht]
\centering{%
\includegraphics[scale=1]{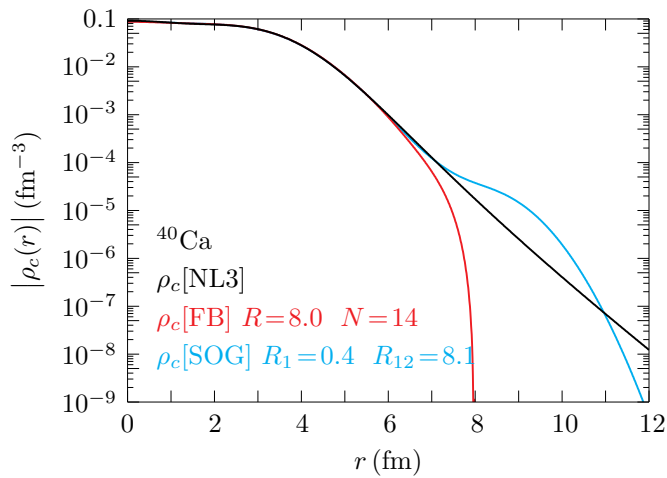}
}
\caption{The charge density of $^{40}$Ca. The red and blue curves are obtained by the Fourier-Bessel(FB)-
and sum-of Gaussians(SOG)-analyses of electron-scattering data, respectively.
The black one is calculated using
the relativistic mean-field model, NL3. The FB- and SOG-coefficients are taken from Ref.\cite{vries}.
In the figure,  $R=8$ fm indicates the cut-off parameter, and $N=14$ the number of the coefficients
in FB-series expansion, while the values of $R_1$ and $R_{12}$ stand for the Gaussian-parameters
in Ref.\cite{vries}.
For details, see the text.}
\label{cd-Ca40}
\end{figure}

\begin{figure}[ht]
\centering{%
\includegraphics[scale=1]{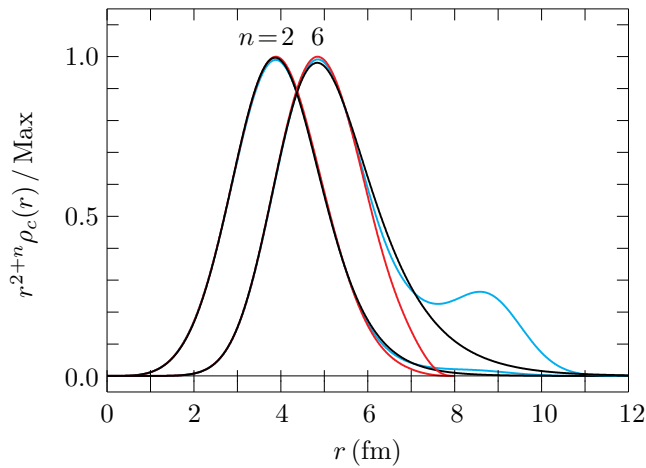}
}
\caption{The charge density multiplied $r^{2+n}$ in $^{40}$Ca.
The charge density is obtained by the Fourier-Bessel(FB)- and
sum-of Gaussians(SOG)-analyses of electron scattering data, as in Fig \ref{cd-Ca40}.
The curves are normalized by the maximum value(Max) of each function. 
For details, see the text.}
\label{rn-Ca40}
\end{figure}

It is well known that for analyzing experimental cross section in electron scattering,
there are methods which are called  ``model-independent'' ways.
The one is Fourier-Bessel(FB)-method\cite{fb},
and the other sum-of-Gaussians(SOG)-method\cite{sog}. In these cases, however,
``model-independent'' means the analysis to provide
charge distribution which fits the experimental cross section
without regard to its special functional form like Fermi-type\cite{fl}.
In fact, the FB-method can not use the infinite series for the analysis, while the SOG-one
assumes a finite number of the  Gaussian distribution of the fractional charge in the nucleus.
Strictly speaking, these analyses imply that if experimental data to be analyzed by FB-method are
given by different momentum-transfer regions from those by the SOG-one,
then the obtained charge distributions may be different from each other.
In principle, in order to determine uniquely the density distribution from experiment,
an infinite number of data may be required.
Hence, it is necessary to know whether or not the experimental values obtained from these
``model-independent'' analyses are appropriate for the present purpose.

In this Appendix,
the problem of the FB- and SOG-methods is pointed out,
which may be avoidable at present for the study of the moments higher than the fourth.
Fig. \ref{cd-Ca40} shows the charge density
in $^{40}$Ca, using the FB(red curve)- and SOG(blue curve)-parameters listed in Ref.\cite{vries}.
As a reference, the charge density calculated with NL3\cite{nl3} is depicted with the black curve. 
According to Ref.\cite{vries}, the cutoff radius($R$) of the FB-series is taken to be $8$ fm,
while the number of the terms $N=14$. The numbers, $R_1$ and $R_{12}$, indicate examples of
the Gaussian-parameters specified in Ref.\cite{vries}.
Fig. \ref{rn-Ca40} shows the charge density of Fig. \ref{cd-Ca40} multiplied by $r^{2+n}(n=2, 6)$,
where the curves are normalized by the maximum value(Max) of each function.

With respect to the values of $R^2_c$ and $Q^4_c$ of the charge distribution,
there is not a large difference between the two analyses, as listed in Table \ref{table_nfs}.
The SOG-values are obtained with the upper limit of the integral for the moments to be $20$ fm.
Ref.\cite{vries} listed the same values of $R_c$ for $^{40}$Ca,
as $3.450(10)$ and $3.480(3)$ fm in FB- and SOG-analyses, respectively.
The values in the NL3 model\cite{nl3} and those of $\avr{r^8}_c$ are also listed
in Table \ref{table_nfs}, as a reference. 

\begin{figure}[ht]
\centering{%
\includegraphics[scale=1]{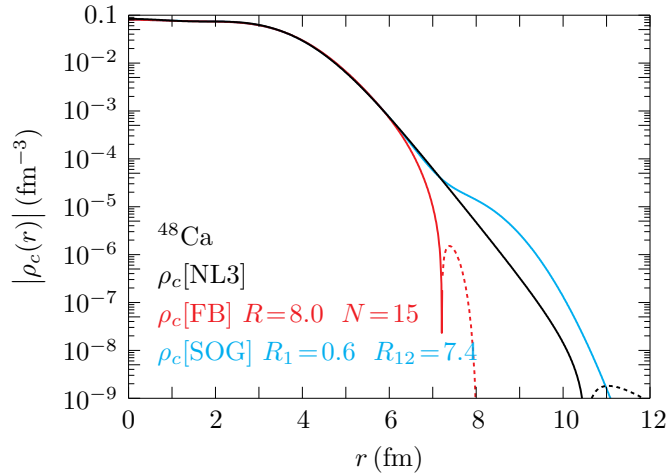}
}
\caption{The charge density of $^{48}$Ca.
The figure is depicted in the same way as Fig. \ref{cd-Ca40},
but the dashed parts of red and black curves indicate the negative densities.
The FB- and SOG-coefficients are taken from Ref.\cite{vries}.
For details, see the text.}
\label{cd-Ca48}
\end{figure}

\begin{figure}[ht]
\centering{%
\includegraphics[scale=1]{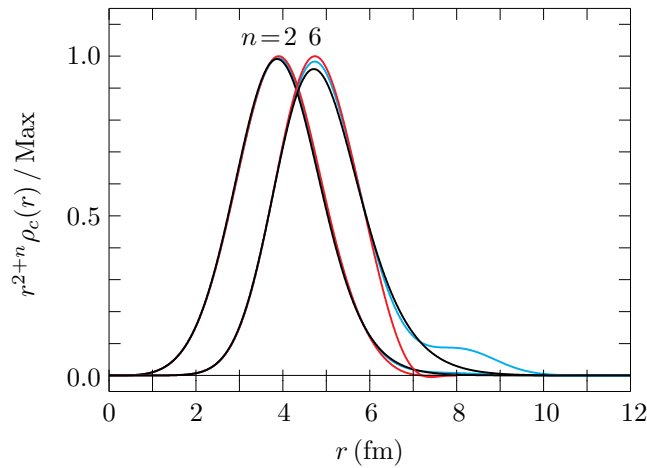}
}
\caption{The charge density multiplied $r^{2+n}$ in $^{48}$Ca.
The figure is depicted in the same way as Fig. \ref{rn-Ca40}. 
For details, see the text.}
\label{rn-Ca48}
\end{figure}

Figures \ref{cd-Ca48} and \ref{rn-Ca48} show the charge density and the one multiplied
by $r^{2+n}$ in $^{48}$Ca, respectively, in the same way as the previous figures for $^{40}$Ca.
The dashed parts of FB(red)- and NL3(black)-curves in Fig. \ref{cd-Ca48} indicate
the negative densities. 
The obtained values of $R_c$ for $^{48}$Ca are listed in Table \ref{table_nfs}, which are the same
as those in Ref.\cite{vries}, as $3.451(9)$ and $3.460(0)$ fm
by the FB- and SOG-analyses, respectively.

Even taking account of experimental errors in the parentheses,
in both $^{40}$Ca and $^{48}$Ca, the values of $R^2_c$ by the SOG-method are a little larger than
than those by the FB-one. Their ratio is listed in the last column of Table \ref{table_nfs}.
In the present paper, the experimental and model-dependent errors
in the FB- and SOG-analyses were not estimated, because there are not enough data
in Ref.\cite{vries} for their evaluations.   
The reason why the SOG-values are larger than the FB-values
is easily seen in Figs. \ref{cd-Ca40} to \ref{rn-Ca48}.
On the one hand, in the SOG-result, there is non-negligible contribution from the tail of
the charge density
to the $R^2_c$, but its contribution depends on the unphysical bump around $r\ge 8$ fm.
On the other hand, in the FB-one, the cutoff parameter, $R=8$ fm,
dominates the shape of the density-tail.
In particular, in $^{48}$Ca, the negative density appears owing to series of the spherical
Bessel functions, whereas that of NL3 around $r=11$ fm is due to the neutron charge density. 
The SOG-components are not negative by definition.

\begin{figure}[ht]
\centering{%
\includegraphics[scale=1]{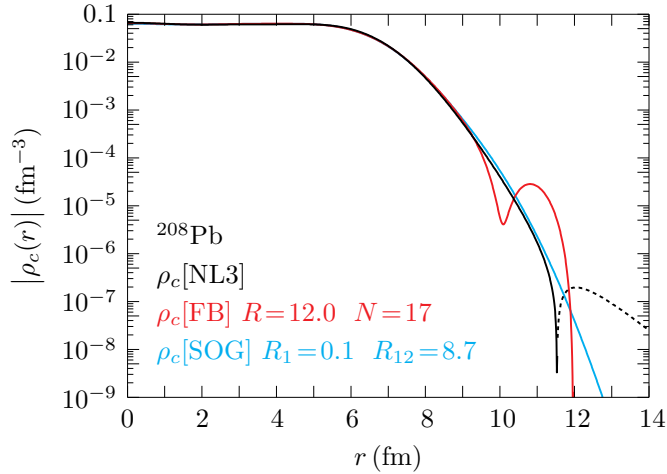}
}
\caption{The charge density of $^{208}$Pb.
The figure is depicted in the same way as Fig. \ref{cd-Ca40},
but the dashed parts of red and black curves indicate the negative densities.
The FB- and SOG-coefficients are taken from Ref.\cite{vries}.
For details, see the text.}
\label{cd-Pb208}
\end{figure}

\begin{figure}[ht]
\centering{%
\includegraphics[scale=1]{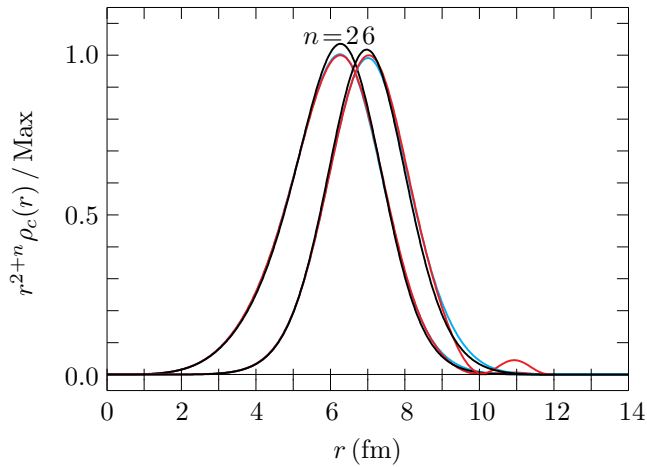}
}
\caption{The charge density multiplied $r^{2+n}$ in $^{208}$Pb.
The figure is depicted in the same way as Fig. \ref{rn-Ca40}. 
For details, see the text.}
\label{rn-Pb208}
\end{figure}

Figures \ref{cd-Pb208} and \ref{rn-Pb208}
are the same as Figs. \ref{cd-Ca40} and \ref{rn-Ca40}, but for $^{208}$Pb.
Ref.\cite{vries} listed in the FB-method the two different values of $R_c$ in $^{208}$Pb
for the two corresponding sets of experimental data. The one was obtained with $R=12.0$ fm, while
the other $R=11.0$ fm. The former, which is used in the present paper, has $17$ coefficients of
FB-series, yielding $R_c=5.503(2)$. The latter has $13$ ones and provides $R_c=5.499(1)$.

In the FB-analysis,
the oscillation of the charge density is observed, whereas in the SOG-one,
the density decreases monotonically with $r$ up to $20$ fm. 
As a results, the ratio of the FB-value to the SOG-one
is $1.0000$ for $R_c$, $Q_c$ and $H_c$ in $^{208}$Pb, as expected from Fig. \ref{rn-Pb208}.
The obtained values are $Q_c=5.851$ and $H_c=6.128$ fm,
and  the value of $R_c$ is the same as $5.503$ fm listed in Ref.\cite{vries}.

Calculations of $R_c$ in an accuracy within $1\%$\cite{hargen} 
should take care of the problems of the density-tails in the FB- and SOG-analyses,
but a few $\%$ difference is not essential for the present discussions.
In the discussions on $H^6_c$ in the present paper, however, the difference between
the FB- and SOG-values are not disregarded, because of the two reasons.
The one is  that the cutoff radius $R=8$ fm for $^{40}$Ca and $^{48}$Ca
in the FB-analyses seems not to be enough for covering most of the region
where the function $r^8\rho_c(r)$ is finite, as seen in Figs. \ref{rn-Ca40}
and \ref{rn-Ca48}.
In the FB-analysis, the value of the $n$th moment is proportional to $R^n$.
In SOG-calculations, the cut-off radius is required to be at least $12$ fm
for the enough convergence of the integral on $H^6_c$.
The other reason is that
the unphysical negative density in FB-distribution
and the bump in SOG-one contribute to the values of $H^6_c$,
as seen in Figs. \ref{rn-Ca40}, \ref{rn-Ca48} and \ref{rn-Pb208}.
Since there are not enough experimental data to improve the results
of Ref.\cite{vries} at present,
the LSA of $H^6_c$ has been performed by using  the both FB- and SOG-values,
but in \S \ref{lsa} the LSA has been shown by using the values of the SOG-method.
The results with the FB-values will be summarized
in \S \ref{sum}, together with those of the SOG-one.
It is noted that Table \ref{table_nfs} shows the worst ratio of
the FB-value to the SOG-one is $0.9606$ for $H_c$ in $^{40}$Ca 
among the moments discussed in the present paper.\footnote[8]{The oscillations 
in the densities obtained by the FB- and SOG-methods have been pointed out
in Ref.\cite{xa} also.}


\begin{thebibliography}{99}
\bibitem{thi} M. Thiel et al., J. Phys. G : Nucl. Part. Phys. \textbf{46}, 093003 (2019).
\bibitem{bd} J. D. Bjorken and S. D. Drell, Relativistic quantum mechanics
	(McGraw Hill Book Company, 1964).	  
\bibitem{deforest}T. deForest and J. D. Walecka, Adv. Phys. \textbf{15}, 1 (1966).
\bibitem{vries}H. De Vries C. W. De Jager and C. De Vries,
	 Atom. Data Nucl.Data Tabl. \textbf{36}, 495 (1987).
\bibitem{ks1} H.kurasawa and T. Suzuki, Prog. Theor. Exp. Phys. \textbf{2019},
	113D01 (2019).
\bibitem{kss} H. Kurasawa, T. Suda and T. Suzuki, Prog. Theor. Exp. Phys. \textbf{2021},
	013D02(2021).
\bibitem{bm} A. Bohr and B. R. Mottelson, Nuclear structure, vol.1
	(World Scientific Publishing Co. Pte. Ltd., 1998).
\bibitem{stone} J. R. Stone et al., Phys.Rev. \textbf{C68}, 034324 (2003).
\bibitem{nl3} G. A. Lalazissis, J. K\"{o}ning and P. Ring, Phys. Rev. \textbf{C55}, 540 (1997).
\bibitem{emrich} H. J. Emrich, PhD thesis,
	Johannes-Gutenberg-Universit$\ddot{a}$t, Mainz,1983.	
\bibitem{nlsh} M. M. Shama, M. A. Nagarajan, and P. Ring, Phys. Lett. \textbf{B312}, 377 (1993).
\bibitem{sly4} E. Chabanat, et al., Nucl. Phys. \textbf{A635}, 231 (1998).
\bibitem{ks0} H. Kurasawa and T. Suzuki, Phys. Rev. \textbf{C62}, 054303 (2000).
\bibitem{vret} D. Vretenar, T. Nik\u{s}i\'{c} and P. Ring, Phys.Rev. \textbf{C68}, 024310 (2003). 

\bibitem{sw1} B. D. Serot and J. D. Walecka, Int. Jour. Mod. Phys. \textbf{E6}, 515 (1997). 
\bibitem{nl1} P. G. Reinhard et al., Z. Phys \textbf{A323}, 13 (1986).
\bibitem{nlz} M. Rufa et al., Phys. Rev. \textbf{C38}, 390 (1988). 
\bibitem{nls} P. G. Reinhard, Z. Phys. \textbf{A329}, 257(1988).
\bibitem{nl32} G. A. Lalazissis, J. K\"{o}nig, and P. Ring, Phys. Rev. \textbf{C55}, 540 (1997).   
\bibitem{tm1} Y. Sugahara and H. Toki, Nucl. Phys. \textbf{A579}, 557 (1994).
\bibitem{fsu}B. G. Todd-Rutel and J. Piekarewicz, Phys. Rev. Lett. \textbf{95}, 122501(2005).
\bibitem{sk1} D. Vautherin and D. M. Brink, Phys. Rev. \textbf{C5}, 626 (1972).
\bibitem{sk3} M. Beiner et al., Nucl. Phys.\textbf{A238}, 29 (1975).
\bibitem{skm} J. Bartel et al., Nucl. Phys. \textbf{A386}, 79 (1982).
\bibitem{st6} E. Chabanat et al., Nucl. Phys. \textbf{A627}, 710 (1997)	
\bibitem{sg2} N. V. Giai and H. Sagawa, Phys. Lett. \textbf{B106}, 379 (1981).
\bibitem{ska} H. S. K\"{o}hler, Nucl. Phys. \textbf{A258}, 301 (1976).


\bibitem{toshio}T. Suzuki, Prog. Theor. Exp. Phys. \textbf{2022}, 063D01 (2022).
\bibitem{ks2}H. Kurasawa and T. Suzuki, Prog. Theor. Exp. Phys. \textbf{2022}, 023D03 (2022).
\bibitem{roca} X. Roca-Mazza, M. Centelles, X. Vi\~{n}as and M. Warda,
	Phys. Rev. Lett., \textbf{106}, 252501 (2011).	

\bibitem{thouless} D. J. Thouless, Nucl. Phys. \textbf{22}, 78 (1961).
\bibitem{tsrpa} T. Suzuki, Ann. de Phys. Fr., \textbf{9}, 535 (1984). 
\bibitem{ksrpa2} H. Kurasawa and T. Suzuki, Prog. Theor. Exp. Phys., \textbf{2013}, 043D04 (2013).  
\bibitem{ksrpa3} H. Kurasawa and T. Suzuki, Prog. Theor. Exp. Phys., \textbf{2015}, 113D02 (2015). 

\bibitem{bethe} H. A. Bethe, Phys. Rev. \textbf{103}, 1353 (1956). 
\bibitem{weisskopf} V. F. Weisskopf, Nucl. Phys. \textbf{3}, 423 (1957).	
\bibitem{hvh} N. M. Hugenholtz and L. Van Hove, Physica \textbf{24}, 363 (1958).
\bibitem{jlab1} S. Abrahamyan et al., Phys. Rev. Lett. \textbf{108}, 112502 (2012).
\bibitem{jlab2} D. Adhikari et al., Phys. Rev Lett. \textbf{126}, 172502 (2021).
\bibitem{jlab3} D. Adhikari et al., Phys. Rev Lett. \textbf{129}, 042501 (2022).

	

\bibitem{frosch}R. F. Frosch, et al., Phys. Rev. \textbf{174}, 1380 (1968). 

\bibitem{hs}C. J. Horowitz and B. D. Serot, Nucl. Phys. \textbf{A368}, 503 (1981).

	
\bibitem{fur}J. F. Dawson and R. J. Furnstahl, Phys. Rev. \textbf{C42}, 2009 (1990).
\bibitem{ma}Z. Ma, N. Van Giai, A. Wandelt, D. Vretenar, and P. Ring,
	Nucl. Phys. \textbf{A686}, 173 (2001).
	
\bibitem{ks50} H. Kurasawa and T. Suzuki, Phys. Lett. \textbf{B474}, 262 (2000).
\bibitem{sly42} E. Chabanat, et al., Nucl. Phys. \textbf{A627}, 710 (1997).
\bibitem{fb} B. Dreher, et al. Nucl. Phys. \textbf{A235}, 219 (1974).
\bibitem{sog} I. Sick, Nucl. Phys. \textbf{A218}, 509 (1974).
\bibitem{fl} J. Friedrich and F. Lenz, Nucl. Phys. \textbf{183}, 523 (1972)	

\bibitem{hargen} G. Hargen et al., Nature Phys. \textbf{12}, 186 (2016).
\bibitem{xa} L. Xayavong and Y. Lim, Arxiv-nucl-th, 2303.00945 (2023).

	

	



	








	
	
\end{thebibliography}
\end{document}